\documentclass[11pt,a4paper]{article}
\usepackage{graphicx}
\usepackage{amsmath}
\usepackage{amssymb}
\usepackage{amsthm}
\usepackage{cite}
\usepackage{subfigure}

\newcommand{\bm}{\boldsymbol}
\newcommand{\vect}[1]{\mathbf{#1}}
\newcommand{\mat}[1]{\mathbf{#1}}

\newcommand{\E}[1]{{\mathbb{E}}\{#1\}}
\newcommand{\cov}[1]{{\mathrm{cov}}(#1)}

\newcommand{\etr}[1]{{\mathrm{etr}}\left\{#1\right\}}
\newcommand{\Tr}[1]{{\mathrm{Tr}}\{#1\}}

\renewcommand{\det}[1]{|#1|}

\newcommand{\chol}[1]{\mathrm{chol}\left(#1\right)}
\newcommand{\smallsum}{{\Sigma}}

\newcommand{\loss}{\ell}
\newcommand{\lossq}{\ell_{\mSigma=\mSigmat+\vq\vq^{H}}}
\newcommand{\lossger}{\ell_{\text{\tiny{GER}}}}
\newcommand{\lossmpdr}{\loss_{\text{\tiny{PA-MPDR}}}}

\newcommand{\elast}{\vect{e}_{N}}
\newcommand{\vb}{\vect{b}}
\newcommand{\vn}{\vect{n}}

\newcommand{\vq}{\vect{q}}
\newcommand{\vx}{\vect{x}}
\newcommand{\vxbar}{\bar{\vx}}
\newcommand{\vt}{\vect{t}}
\newcommand{\vtbar}{\bar{\vt}}
\newcommand{\vu}{\vect{u}}
\newcommand{\vv}{\vect{v}}
\newcommand{\vw}{\vect{w}}
\newcommand{\wopt}{\vw_{\text{\tiny{opt}}}}
\newcommand{\vzero}{\vect{0}}

\newcommand{\A}{\mat{A}}
\newcommand{\Atilde}{\tilde{\A}}
\newcommand{\B}{\mat{B}}
\newcommand{\Btilde}{\tilde{\B}}
\newcommand{\F}{\mat{F}}
\newcommand{\Ft}{\F_{t}}
\newcommand{\G}{\mat{G}}
\newcommand{\Gt}{\G_{t}}
\newcommand{\I}{\mat{I}}
\newcommand{\eye}[1]{\I_{#1}}

\newcommand{\Q}{\mat{Q}}
\newcommand{\Qv}{\Q_{\mathrm{v}}}

\newcommand{\St}{\mat{S}_{t}}
\newcommand{\U}{\mat{U}}
\newcommand{\V}{\mat{V}}
\newcommand{\Vorth}{\V_{\perp}}
\newcommand{\W}{\mat{W}}
\newcommand{\X}{\mat{X}}
\newcommand{\Xbar}{\bar{\X}}

\newcommand{\Mzero}{\mat{0}}

\newcommand{\veta}{\bm{\eta}}

\newcommand{\mLambda}{\bm{\Lambda}}
\newcommand{\mSigma}{\bm{\Sigma}}
\newcommand{\mSigmat}{\bm{\Sigma}_{t}}
\newcommand{\mOmega}{\bm{\Omega}}
\newcommand{\mPsi}{\bm{\Psi}}
\newcommand{\mPhi}{\bm{\Phi}}

\newcommand{\vCN}[3]{\mathbb{C}\mathcal{N}_{#1}\left(#2,#3\right)}
\newcommand{\mCN}[4]{\mathbb{C}\mathcal{N}_{#1}\left(#2,#3,#4\right)}
\newcommand{\CW}[3]{\mathbb{C}\mathcal{W}_{#1}\left(#2,#3\right)}
\newcommand{\chisquare}[2]{\chi^{2}_{#1}(#2)}
\newcommand{\Cchisquare}[2]{\mathbb{C}\chi^{2}_{#1}(#2)}
\newcommand{\CF}[2]{\mathbb{C}\mathcal{F}_{#1}(#2)}
\newcommand{\betapdf}[2]{\mathcal{B}_{#1}(#2)}
\newcommand{\dist}{\overset{d}{=}}
\newcommand{\approxdist}{\overset{d}{\approx}}

\newcommand{\SNR}{\mathrm{SNR}}

\begin{document}
\title{Analysis of the SNR loss distribution with covariance mismatched training samples}
\author{Olivier Besson \thanks{O. Besson is with ISAE-SUPAERO, Universit\'{e} de Toulouse, 10 avenue Edouard Belin, 31055 Toulouse, France. Email: olivier.besson@isae-supaero.fr}}
\date{31 August 2020}
\maketitle
\begin{abstract}
We analyze the distribution of the signal to noise ratio (SNR) loss at the output of an adaptive filter which is trained with samples that do not share the same covariance matrix as the samples for which the filter is foreseen. Our objective is to find an accurate approximation of the distribution of the SNR loss which has a similar form as in the case of no mismatch. We successively consider the case where the two covariance matrices satisfy the so-called generalized eigenrelation and the case where they are arbitrary. In the former case, this amounts to approximate a central quadratic form in normal variables while the latter case entails approximating a non-central quadratic form in Student distributed variables. In order to obtain the approximate distribution, a Pearson type approach is advocated. A numerical study show that this approximation is rather accurate and enables one to assess, in a straightforward manner, the impact of covariance mismatch. 
\end{abstract}

\section{Problem statement}
Enhancing the reception of a signal of interest (SoI) in the presence of noise by means of a linear filter is an omnipresent issue in many engineering applications, including radar and communications \cite{Scharf91,VanTrees02}.  It has also applications in other fields, for instance in finance with the problem of selecting a mean-variance efficient portfolio \cite{Markowitz52}. The linear filter $\vw$ is usually designed to ensure a unit-gain towards the SoI $\vv$ while minimizing the output power so as to mitigate noise as much as possible. Proceeding this way results in the optimal filter  $\wopt = (\vv^{H} \mSigma^{-1} \vv)^{-1} \mSigma^{-1} \vv$ where  $\mSigma$ stands for the true noise covariance matrix  of the data to be filtered. $\mSigma$ is generally unknown and hence the filter needs to be trained with samples whose actual covariance matrix $\mSigmat$ is hopefully equal to $\mSigma$. Then $\vw = (\vv^{H} \St^{-1} \vv)^{-1} \St^{-1} \vv$ is used in place of $\wopt$ where $\St$ denotes the sample covariance matrix (SCM) of the training samples. In order to assess the performance of $\vw$,  a widely spread measure of efficiency is the SNR loss, i.e., the SNR at the output of $\vw$ divided by the SNR at the output of $\wopt$. The expression of the SNR loss is  given by
\begin{equation}\label{SNRloss}
\loss = \frac{\SNR(\vw)}{\SNR(\wopt)} = \frac{(\vv^H \St^{-1} \vv)^{2}}{(\vv^{H}\mSigma^{-1}\vv)(\vv^{H}\St^{-1}\mSigma\St^{-1}\vv)}.
\end{equation}
Under the Gaussian assumption and assuming that $\mSigmat = \mSigma$, the distribution of $\loss$ was derived, e.g. in \cite{Reed74,Hanumara86,Khatri87} where it was shown that it follows a beta distribution and that it admits the following stochastic representation:
\begin{equation}\label{pdf_SNRloss_nomismatch}
\loss_{\mSigmat=\mSigma} \dist \left[1 + \frac{\chisquare{2(N-1)}{0}}{\chisquare{2(K-N+2)}{0}}\right]^{-1}
\end{equation}
where $\dist$ means ``has the same distribution as'', $\chisquare{\nu}{0}$ denotes the central chi-square distribution with $\nu$ degrees of freedom, $N$ is the size of the observations and $K$ the number of training samples.

However, in practice the training samples may have a covariance matrix different from $\mSigma$. The simplest case is the so-called partially homogeneous environment where $\mSigmat = \gamma \mSigma$. A second common example is the case  where the training samples contain the SoI, i.e., $\mSigmat = \mSigma + P \vv \vv^{H}$. This corresponds to what Van Trees refers to as a minimum power distortionless response (MPDR) scenario \cite{VanTrees02}. It has been thoroughly analyzed in \cite{Boroson80,Monzingo80} where it was showed that the convergence rate of $\vw$ is dramatically degraded and that the SNR loss is significantly larger.  The case where the training samples are contaminated by signal-like components has also been addressed by Gerlach in \cite{Gerlach95} who also considered corruption by outliers in \cite{Gerlach02}.  Rank-one modifications $\mSigmat = \mSigma + \vq \vq^{H}$ have been considered in \cite{Richmond00c,Besson07c,Besson07g}.   The impact of covariance mismatch on the generalized likelihood ratio test, the adaptive matched filter and the adaptive coherence estimator was analyzed in \cite{Blum00,McDonald00,Richmond00b} under some specific assumptions. An important contribution to the analysis of the effects of covariance mismatch is due to Richmond \cite{Richmond00b} who analyzed the SNR loss under the assumption that $\mSigmat$ and $\mSigma$ satisfy the so-called generalized eigenrelation (GER) which states that $\mSigmat^{-1} \vv = \lambda \mSigma^{-1}\vv$. Although initially introduced as a technical assumption that simplifies derivations, it was shown that the GER is physically meaningful \cite{Richmond00b}. Assuming the GER is satisfied, the exact distribution of $\loss$ was obtained as the inverse Laplace transform of the moment generating function and given as a finite sum of functions, see equations(9)-(11) of  \cite{Richmond00b}.  The recent paper \cite{Raghavan19} is the first to address the effect of covariance mismatched training samples for \emph{arbitrary} $\mSigmat$ and $\mSigma$. More precisely,  Raghavan analyzes the performance of the adaptive matched filter (AMF) and provides a representation of the AMF test statistic under both the null and the alternative hypotheses, whatever $\mSigmat$ and $\mSigma$. The approach proposed in \cite{Raghavan19} provides the fundamental tools we will use here to analyze the SNR loss. 

The objective and the contribution of this paper are to provide an approximation of the distribution of $\loss$ of the form
\begin{equation}\label{approx_representation_SNRloss}
\loss_{\mSigmat \neq \mSigma}\approxdist \left[1 + a \frac{\chisquare{\nu}{0}}{\chisquare{\mu}{0}}\right]^{-1}
\end{equation}
where the coefficients $a$, $\nu$ and $\mu$ are to be found. The form in \eqref{approx_representation_SNRloss}  is chosen to resemble that of \eqref{pdf_SNRloss_nomismatch} where $a=1$, $\nu=2(N-1)$ and $\mu=2(K-N+2)$ in the case of no mismatch. The interest of such an approximation is that it is simple, provides a simple and closed-form expression of the probability density function (p.d.f.) of $\loss$ as
\begin{equation}\label{approx_pdf_SNRloss}
p(\loss) \approx \frac{a^{\mu}\Gamma(\nu+\mu)}{\Gamma(\nu)\Gamma(\mu)} \frac{\loss^{\mu-1}(1-\loss)^{\nu-1}}{(1+(a-1)\loss)^{\nu+\mu}}
\end{equation}
Moreover, as will be illustrated later, this approximation turns out to be rather accurate. Finally, it enables one to quickly figure out the impact of covariance mismatch. In the sequel we will first provide a general stochastic representation of $\loss$. Then, we will successively address the case where the GER is satisfied (which includes a partially homogeneous MPDR scenario) and the case of arbitrary $\mSigmat$. The corresponding problems will be respectively that of approximating a central quadratic form in normal variables and that of approximating a non-central quadratic form in Student distributed variables. For both situations a Pearson-like moment approximation will be used. A numerical study will assess the accuracy of this approximation in predicting the distribution of $\loss$ for various types of mismatch.

\vspace*{0.2cm}

\textbf{Notations}: In the paper, column vectors will be denoted as boldface letters, e.g. $\vx$ and matrices will be denoted as capital boldface letters, e.g., $\X$. The identity matrix of size $N$ will be denoted $\eye{N}$. $\det{.}$ and $\etr{.}$ stands for the determinant and the exponential of the trace of a matrix. The Cholesky factor of a positive definite matrix $\mSigma$ is denoted by $\G=\chol{\mSigma}$. It is the unique lower triangular matrix with real-valued positive diagonal elements such that $\G\G^{H}=\mSigma$. A $N$-length vector $\vx$ follows a complex multivariate Gaussian distribution, which we denote as $\vx \dist \vCN{N}{\vxbar}{\mPhi}$ if its p.d.f. can be written as $p(\vx) = \pi^{-N} \det{\mPhi} \etr{-(\vx-\vxbar)^{H} \mPhi^{-1} (\vx-\vxbar)}$.  A $N \times K$ matrix $\X$ is said to  follow a complex matrix-variate Gaussian distribution, which we denote as $\X \dist \mCN{N,K}{\Xbar}{\mPhi}{\mPsi}$ if its density is $p(\X) = \pi^{-NK} \det{\mPhi}^{-K} \det{\mPsi}^{-N}\etr{-(\X-\Xbar)^{H}\mPhi^{-1}(\X-\Xbar)\mPsi^{-1}}$. If $\X \dist \mCN{N,K}{\Mzero}{\mSigma}{\eye{K}}$ then $\W = \X \X^{H} \dist \CW{N}{K}{\mSigma}$ follows a complex Wishart distribution with parameters $K$ and $\mSigma$. The complex chi-square distribution with $p$ degrees of freedom and non centrality parameter $\delta$ is denoted $\Cchisquare{p}{\delta}$. It is the distribution of $\vx^{H}\vx$ when $\vx \dist \vCN{p}{\vxbar}{\eye{\nu}}$ and $\delta =  \vxbar^{H}\vxbar$. If $\chisquare{p}{\delta}$ is the usual (real) chi-square, one has $\Cchisquare{p}{\delta} \dist 0.5 \chisquare{2p}{2\delta}$. Let $U \dist \Cchisquare{p}{\delta}$ and $V \dist \Cchisquare{q}{0}$. Then $F = U/V$ follows a complex F-distribution, which is noted as  $F \dist \CF{p,q}{\delta}$. $B=(1+F)^{-1} \dist \betapdf{p,q}{\delta}$.  When $\delta=0$, the probability density function of $B$ is given by $p_{B}(\beta) = \frac{\beta^{q-1}(1-\beta)^{p-1}}{B(p,q)}$ where $B(p,q) = \frac{\Gamma(p)\Gamma(q)}{\Gamma(p+q)}$.  For a random variable $Q$, we will note $\E{Q}$ its mean value, $\mu_{n}(Q)$ and $\kappa_{n}(Q)$ its $n$-th central moment and cumulant. 

\section{Statistical representation of the SNR loss}
In this section, we present a general statistical representation of the SNR loss. The procedure involves a first step of whitening followed by a rotation to bring the signal of interest along a vector composed of zeroes except for its last component. Then we resort to results about partitioned Wishart matrices. The technical derivations leading to the representation in equation \eqref{SNRloss_vs_Omega} bear much resemblance with the derivations in \cite{Raghavan19} where the probability of false alarm of the AMF is derived. Most of the variables to be defined next, e.g., $\Q$, $\mOmega$ have equivalent counterparts in \cite{Raghavan19}. It should also be mentioned that a representation of the SNR loss is provided in \cite{Raghavan19}, see its equation (26). 

Let $\Gt=\chol{\mSigmat}$ and note that the SCM of the training samples $\St \dist \Gt \W \Gt^{H} $ where $\W  \dist \CW{N}{K}{\eye{N}}$ and $\St \dist  \Gt \Q \W \Q^{H} \Gt^{H} $ for any unitary matrix $\Q$. Let $\Vorth$ be a semi-unitary matrix orthogonal to $\vv$, i.e., $\Vorth^{H}\Vorth =\eye{N-1}$ and $\Vorth^{H}\vv=\vzero$ and let us assume without loss of generality that $\left\| \vv  \right\|=1$. Also, let $\Ft = \chol{\Vorth^{H} \mSigmat \Vorth}$ and let us construct $\Q = \begin{bmatrix} \Gt^{H} \Vorth \Ft^{-H} & (\vv^{H}\mSigmat^{-1}\vv)^{1/2} \Gt^{-1} \vv \end{bmatrix}$ such that $\Q^{H}\Gt^{-1}\vv=(\vv^{H}\mSigmat^{-1}\vv)^{1/2} \elast$ where $\elast = \begin{bmatrix} \vzero^{T} & 1 \end{bmatrix}^{T}$. Then
\begin{align}
\loss &= \frac{(\vv^H \St^{-1} \vv)^{2}}{(\vv^{H}\mSigma^{-1}\vv)(\vv^{H}\St^{-1}\mSigma\St^{-1}\vv)} \nonumber \\
&\dist  \frac{(\vv^H \Gt^{-H} \Q \W^{-1} \Q^{H} \Gt^{-1} \vv)^{2}}{(\vv^{H}\mSigma^{-1}\vv)(\vv^{H}\Gt^{-H} \Q \W^{-1} \Q^{H} \Gt^{-1}\mSigma \Gt^{-H} \Q \W^{-1} \Q^{H} \Gt^{-1} \vv)} \nonumber \\
&= \frac{\vv^{H}\mSigmat^{-1}\vv}{\vv^{H}\mSigma^{-1}\vv} \frac{(\elast^{H}\W^{-1}\elast)^{2}}{\elast^{H}\W^{-1}\mOmega \W^{-1}\elast}
\end{align}
where
\begin{align}\label{Omega}
\mOmega &= \Q^{H} \Gt^{-1}\mSigma \Gt^{-H} \Q  \nonumber \\
&= \begin{pmatrix}  \Ft^{-1} \Vorth^{H} \mSigma \Vorth \Ft^{-H} & \frac{\Ft^{-1} \Vorth^{H} \mSigma\mSigmat^{-1}\vv}{(\vv^{H}\mSigmat^{-1}\vv)^{1/2}}  \\
\frac{\vv^{H} \mSigmat^{-1}\mSigma\Vorth \Ft^{-H}}{(\vv^{H}\mSigmat^{-1}\vv)^{1/2}}  & \frac{\vv^{H} \mSigmat^{-1}\mSigma \mSigmat^{-1}\vv}{\vv^{H}\mSigmat^{-1}\vv} \end{pmatrix} \nonumber \\
&=\begin{pmatrix} \mOmega_{11} & \mOmega_{12} \\ \mOmega_{21} & \Omega_{22} \end{pmatrix}
\end{align}
Note that, when $\mSigmat=\mSigma$, $\mOmega=\eye{N}$. Next, if we partition $\W$ as in \eqref{Omega}, we get
\begin{align}
\W^{-1}\elast &=  \begin{pmatrix} \W_{1.2}^{-1} & -\W_{11}^{-1} \W_{12 } W_{2.1}^{-1} \\ - W_{2.1}^{-1} \W_{21} \W_{11}^{-1} & W_{2.1}^{-1} \end{pmatrix}  \begin{pmatrix} \vzero \\ 1 \end{pmatrix} \nonumber \\
& = W_{2.1}^{-1} \begin{pmatrix}  -\vt_{12 }  \\ 1 \end{pmatrix}
\end{align}
where 
\begin{align*}
W_{2.1}&=W_{22} - \W_{21}\W_{11}^{-1} \W_{12 } \\
\vt_{12}& = \W_{11}^{-1} \W_{12 }
\end{align*}
The SNR loss can thus be written as
\begin{equation}
\loss = \frac{\vv^{H}\mSigmat^{-1}\vv}{\vv^{H}\mSigma^{-1}\vv}  \, \frac{1}{\begin{bmatrix} -\vt_{12}^{H} & 1 \end{bmatrix} \mOmega \begin{bmatrix} -\vt_{12} \\ 1 \end{bmatrix}  }
\end{equation}
However,
\begin{align}
\begin{bmatrix} -\vt_{12}^{H} & 1 \end{bmatrix} \mOmega \begin{bmatrix} -\vt_{12} \\ 1 \end{bmatrix}  &= \vt_{12}^{H} \mOmega_{11} \vt_{12} - \mOmega_{21}\vt_{12} - \vt_{12}^{H} \mOmega_{12} + \Omega_{22} \nonumber \\
&= (\vt_{12}-\mOmega_{11}^{-1}\mOmega_{12})^{H} \mOmega_{11} (\vt_{12}-\mOmega_{11}^{-1}\mOmega_{12}) + \Omega_{2.1}
\end{align}
which, along with the readily verified fact that $\Omega_{2.1} =  \frac{\vv^{H}\mSigmat^{-1}\vv}{\vv^{H}\mSigma^{-1}\vv}$ yields the following compact expression
\begin{equation}\label{SNRloss_vs_Omega}
\loss = \left[ 1+ \Omega_{2.1}^{-1}  (\vt_{12}-\vtbar_{12})^{H} \mOmega_{11} (\vt_{12}-\vtbar_{12}) \right]^{-1}
\end{equation}
with $\vtbar_{12}=\mOmega_{11}^{-1}\mOmega_{12}$. Now from \cite{Khatri87} $\vt_{12}$ follows a complex multivariate Student distribution and can be represented as
\begin{equation}
\vt_{12} \dist \frac{\vn_{12}}{\sqrt{V_{12}}} \dist \frac{\vCN{N-1}{\vzero}{\eye{N-1}}}{\sqrt{\Cchisquare{K-N+2}{0}}}
\end{equation}
At this stage it is clear that the value of $\vtbar_{12}$ plays a crucial rule. If $\vtbar_{12}=\vzero$, then
\begin{equation}
\vt_{12}^{H} \mOmega_{11} \vt_{12} = V_{12}^{-1} \vn_{12}^{H} \mOmega_{11} \vn_{12}
\end{equation}
and we need to find (or approximate) the distribution of a quadratic form in normal variables, a problem that has been already addressed in the literature. On the other hand when $\vtbar_{12} \neq \vzero$ we need to handle a non central quadratic form in Student distributed variables, a much more challenging problem.  
Before proceeding let us examine what $\vtbar_{12}=\vzero$ means. We have
\begin{align}
\vtbar_{12}=\vzero &\Leftrightarrow \mOmega_{12} = \vzero \nonumber \\
&\Leftrightarrow \Vorth^{H} \mSigma\mSigmat^{-1}\vv = \vzero \nonumber \\
&\Leftrightarrow  \mSigma\mSigmat^{-1}\vv = \lambda \vv \nonumber \\
&\Leftrightarrow  \mSigmat^{-1}\vv = \lambda \mSigma^{-1}\vv 
\end{align}
which is the GER. Therefore, the GER, when it holds, allows for a significant simplification of the problem. We will thus firstly examine this case before addressing the general case. Prior to that, we rewrite \eqref{SNRloss_vs_Omega} in a more suitable form. Let $\mOmega_{11} = \U \mLambda \U^{H}$ denote the eigenvalue decomposition of $\mOmega_{11}$. Then
\begin{align}\label{Q_Student}
Q &= (\vt_{12}-\vtbar_{12})^{H} \mOmega_{11} (\vt_{12}-\vtbar_{12}) \nonumber \\
&= V_{12}^{-1} (\vn_{12}-V_{12}^{1/2}\vtbar_{12})^{H} \U \mLambda \U^{H} (\vn_{12}-V_{12}^{1/2}\vtbar_{12})  \nonumber \\
&= V_{12}^{-1} (\U^{H}\vn_{12}-V_{12}^{1/2}\U^{H}\vtbar_{12})^{H} \mLambda  (\U^{H}\vn_{12}-V_{12}^{1/2}\U^{H}\vtbar_{12})  \nonumber \\
&= V_{12}^{-1} \sum_{i=1}^{N-1} \lambda_{i} \left| (\U^{H}\vn_{12}-V_{12}^{1/2}\U^{H}\vtbar_{12})_{i}\right|^{2} \nonumber \\
&\dist V_{12}^{-1} \sum_{i=1}^{N-1} \lambda_{i} \Cchisquare{1}{V_{12} |\vu_{i}^{H}\vtbar_{12}|^{2}} \nonumber \\
& \dist [2V_{12}]^{-1} \sum_{i=1}^{N-1} \lambda_{i} \chisquare{2}{2V_{12}  | \vu_{i}^{H}\vtbar_{12}|^{2}} \nonumber \\
& \dist V^{-1} \sum_{i=1}^{N-1} \lambda_{i} \chisquare{2}{V  \delta_{i}} 
\end{align}
where $V =2V_{12} \dist \chisquare{2(K-N+2)}{0}$ and $\delta_{i}=| \vu_{i}^{H}\vtbar_{12}|^{2} = | \vu_{i}^{H}\mOmega_{11}^{-1}\mOmega_{12}|^{2} = |\lambda_{i}^{-1}\vu_{i}^{H}\mOmega_{12}|^{2} $. Therefore, the SNR loss admits the following representation
\begin{equation}\label{SNRloss_vs_lambda}
\loss \dist \left[ 1+ \Omega_{2.1}^{-1} \frac{\sum_{i=1}^{N-1} \lambda_{i} \chisquare{2}{V  \delta_{i}}}{V} \right]^{-1}
\end{equation}
with $V  \dist \chisquare{2(K-N+2)}{0}$. When $\mSigmat=\mSigma$, $\mOmega=\eye{N}$, $\lambda_{i}=1$, $\delta_{i}=0$ and we recover \eqref{pdf_SNRloss_nomismatch}. We will now successively investigate approximations of this distribution, first by assuming that the GER holds, next for arbitrary $\mSigma$ and $\mSigmat$. 

\section{Analysis of the SNR loss under the generalized eigenrelation}
In this section we assume that the GER is satisfied, i.e., $\mSigmat^{-1}\vv = \lambda \mSigma^{-1}\vv $. We first examine the case of a MPDR scenario, for which the exact distribution of the SNR loss will be given, then the case of a surprise or undernulled interference, and finally the general case.
\subsection{The case $\mSigmat = \gamma \mSigma + P \vv \vv^{H}$}
We first consider a \emph{partially homogeneous MPDR} scenario where the training samples contain the signal of interest, and the environment may be partially homogeneous, i.e., the noise power differs between the training samples and the samples to be processed. This case was addressed in \cite{Boroson80,Monzingo80} with $\gamma=1$, yet no expression of the SNR loss p.d.f. was given, rather surprisingly. Since $\Vorth^{H} \mSigmat \Vorth = \gamma \Vorth^{H} \mSigma \Vorth$, it follows that $\mOmega_{11}=\gamma^{-1} \eye{N-1}$. Moreover, 
\begin{equation}
\mSigmat^{-1}\vv = \frac{\gamma^{-1}}{1+\gamma^{-1}P \vv^{H}\mSigma^{-1}\vv} \mSigma^{-1}\vv
\end{equation}
which implies that
\begin{equation}\label{Omega_MPDR}
\mOmega = \begin{pmatrix}\gamma^{-1} \eye{N-1} & \Mzero \\ \vzero &  \frac{\gamma^{-1}}{1+\gamma^{-1}P \vv^{H}\mSigma^{-1}\vv} \end{pmatrix}
\end{equation}
Therefore $\lambda_{i}=\gamma^{-1}$, $\Omega_{2.1}=\gamma^{-1}(1+\gamma^{-1}P \vv^{H}\mSigma^{-1}\vv)^{-1}$ and the SNR loss in \eqref{SNRloss_vs_lambda}  can be rewritten as
\begin{equation}\label{SNRloss_MPDR}
\lossmpdr \dist \left[1 + a \frac{\chisquare{2(N-1)}{0}}{\chisquare{2(K-N+2)}{0}}\right]^{-1}
\end{equation}
with $a=1+\gamma^{-1}P \vv^{H}\mSigma^{-1}\vv$.  Its \emph{exact p.d.f.} is given by the right-hand side of \eqref{approx_pdf_SNRloss} with $\nu=2(N-1)$ and $\mu=2(K-N+2)$.
When no SoI is present in the training samples, $a=1$ and one recovers the usual beta distribution. Note that, as $a$ increases, the detrimental effect of the SoI presence is more pronounced, i.e., the distribution of the SNR loss is moved to smaller values. Interestingly enough, a partially homogeneous noise environment can amplify [$\gamma < 1$] or, on the contrary, attenuate [$\gamma > 1$] the deleterious impact of the SoI presence in the training samples.

\begin{figure}[h]
	\centering
	\includegraphics[width=8.5cm]{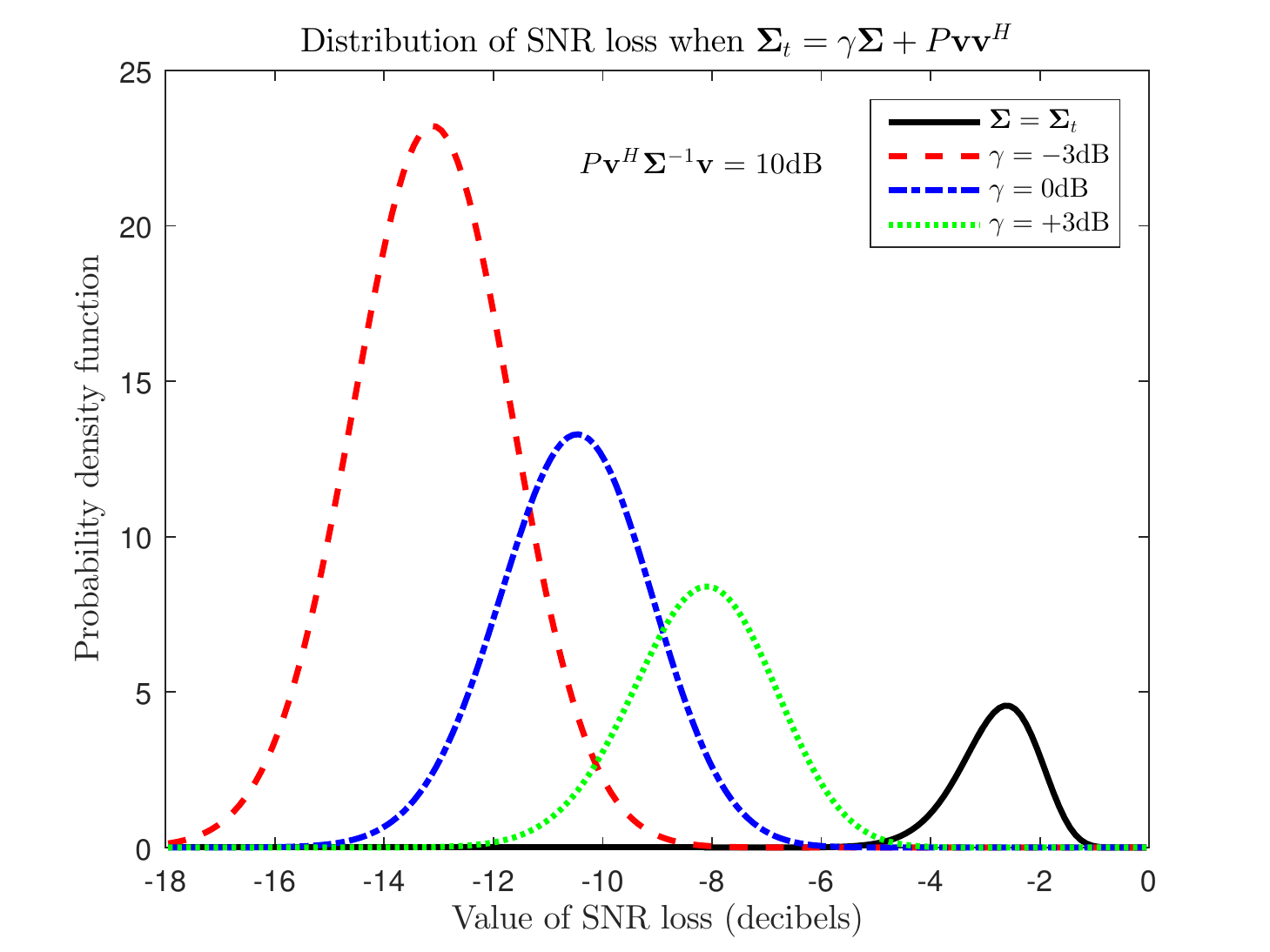}
	\caption{Distribution of the SNR loss in the presence of the SoI in the training samples versus $\gamma$.}
	\label{fig:pdf_snrloss_MPDR_K=32}
\end{figure}

In order to illustrate the impact of the SoI presence in the training samples, we consider, as in the rest of the paper, a scenario with a $N=16$ elements uniform linear array with half wavelength spacing. The noise consists of thermal (white Gaussian) noise and $3$ interfering signals located at $-12^{\circ}$, $9^{\circ}$ and $25^{\circ}$ (measured with respect to the normal of the array), with respective powers $35$dB, $25$dB and $30$dB above thermal noise power. The SoI has direction of arrival $0^{\circ}$ and its power $P$ is such that $P\vv^{H}\mSigma^{-1}\vv=10$dB. The number of training samples is fixed to $K=2N$. Figure \ref{fig:pdf_snrloss_MPDR_K=32} displays the distribution of the SNR loss for different values of $\gamma$. For comparison purposes, the p.d.f. of $\loss$ in the absence of mismatch is also plotted. As can be seen, the noise power in-homogeneity between the training samples and the processed samples has a significant impact, even for rather small variations (here $\pm 3$dB) of $\gamma$, whether in a better or in a worse way.

\subsection{The case $\mSigma = \mSigmat +  \vq \vq^{H}$}
\begin{figure}[h]
	\centering
	\includegraphics[width=8.5cm]{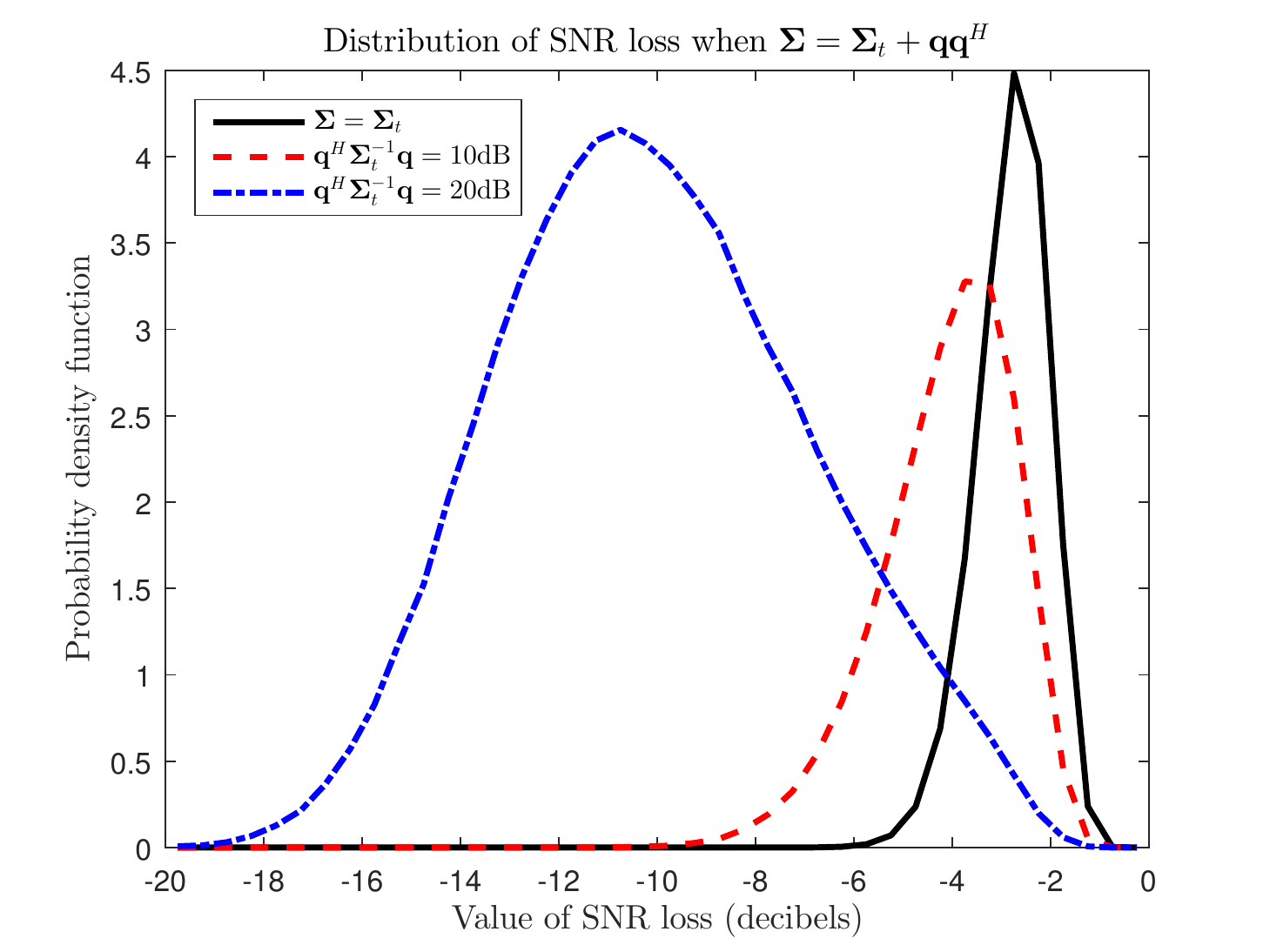}
	\caption{Distribution of the SNR loss in the presence of a surprise interference located at $14^{\circ}$.}
	\label{fig:pdf_snrloss_ger_undernulled_angle_q=14_K=32}
\end{figure}

We now consider the case of a surprise or undernulled interference \cite{Richmond00c} which is present in the test data, yet not accounted for in the training samples and we assume that its signature is proportional to a vector $\vq$. For instance, in the example to be presented in Figure \ref{fig:pdf_snrloss_ger_undernulled_angle_q=14_K=32}, $\vq$ corresponds to the steering vector of an interfering signal impinging on the array. The GER, to be satisfied, requires that $\vq^{H} \mSigma^{-1}\vv=0$, which implies that $\mSigmat^{-1}\vv = \mSigma^{-1}\vv $ and hence $\Omega_{2.1}=1$. Since $\Vorth^{H} \mSigma \Vorth =  \Vorth^{H} \mSigmat \Vorth + \Vorth^{H} \vq \vq^{H} \Vorth$, it follows that
\begin{equation}\label{Omega_nulled}
\mOmega_{11} =  \Ft^{-1} \Vorth^{H} \mSigma \Vorth \Ft^{-H} = \eye{N-1} + \vu \vu^{H}
\end{equation}
where $\vu = \Ft^{-1} \Vorth^{H} \vq$. Therefore $\mOmega_{11}$ has $2$ distinct eigenvalues: $\lambda_{1}=1$ with multiplicity $N-2$ and $\lambda_{2} = 1+\vu^{H}\vu$ with multiplicity $1$. Now, it is possible to show that $\vu^{H}\vu = \vq^{H}\mSigmat^{-1}\vq$ and hence $\lambda_{2} = 1+\vq^{H}\mSigmat^{-1}\vq$. Using \eqref{Omega_nulled} in \eqref{SNRloss_vs_lambda} it follows that
\begin{equation}\label{SNRloss_q}
\lossq \dist \left[1 + \frac{\chisquare{2(N-2)}{0} + (1+\vq^{H}\mSigmat^{-1}\vq) \chisquare{2}{0} }{\chisquare{2(K-N+2)}{0}}\right]^{-1}
\end{equation}
The previous equation provides the exact representation of $\loss$ in this case. It is clear that, in this case, the performance is mostly dictated by $\vu^{H}\vu$: as this quantity increases, or equivalently, as $\vq^{H}\mSigmat^{-1}\vq$ increases, the distribution of the SNR loss is moved towards lower values.  This is illustrated in Figure \ref{fig:pdf_snrloss_ger_undernulled_angle_q=14_K=32} where the data to be processed contains an additional interference located at $14^{\circ}$ with varying power. One can observe again that the SNR loss is degraded because the filter did not learn to cancel this surprise interference.

\subsection{The generic case $\mSigmat^{-1}\vv = \lambda \mSigma^{-1}\vv $}
We do no longer assume a specific relation between $\mSigma$ and $\mSigmat$, simply that they obey the GER. In this case $\vtbar_{21}=\vzero$ and from \eqref{SNRloss_vs_lambda}
\begin{equation}\label{SNRloss_GER}
\lossger \dist \left[ 1+ \Omega_{2.1}^{-1}  \frac{\sum_{i=1}^{N-1} \lambda_{i} \chisquare{2}{0}}{\chisquare{2(K-N+2)}{0}} \right]^{-1}
\end{equation}
Therefore the problem amounts to approximating the distribution of $\smallsum_{i} \lambda_{i} \chisquare{2}{0}$ which is a quadratic form in central normal variables. The problem of approximating the distribution of quadratic form in non-central normal variables is rather well-known, see the book \cite{Mathai92} for a thorough treatment and e.g., \cite{Pearson59,Imhof61,Solomon77,Solomon78,Kuonen99,Liu09,Mohsenipour12,AlNaffouri16,Ramirez19,Ramirez19b} and references therein. One of the most popular approximation is based on Pearson's method \cite{Pearson59} who studied the approximation $\chisquare{\nu}{\delta} \approxdist a_{1}\chisquare{\nu'}{0} + a_{2}$ where the coefficients $a_{1},a_{2},\nu'$ are chosen such that the two random variables have identical mean, standard deviation and skewness, or equivalently identical three first cumulants. Based on this principle, Imohf \cite{Imhof61} derived the best three-moment approximation of $\smallsum_{i} \lambda_{i} \chisquare{h_{i}}{\delta_{i}}$ by $Q_{a} = a_{1}\chisquare{\nu'}{0} + a_{2}$. More precisely, he showed that the solution is given by
\begin{equation}\label{Qa_Imhof}
Q_{a} = \frac{c_{3}}{c_{2}} \chisquare{c_{3}^{2}/c_{2}^{3}}{0} + c_{1} - \frac{c_{2}^{2}}{c_{3}}
\end{equation}
where $c_{s} = [2^{s-1}(s-1)!]^{-1}\kappa_{s}(Q) = \smallsum \lambda_{i}^{s} (h_{i}+s\delta_{i})$.

Coming back to \eqref{SNRloss_GER}, we have $h_{i}=2$, $\delta_{i}=0$ and, since the GER holds, $\Omega_{2.1}=\lambda$, $\mOmega = \begin{pmatrix} \mOmega_{11} & \Mzero \\ \vzero & \lambda \end{pmatrix}$, which implies that
\begin{align}\label{cs_GER}
2^{-1} c_{s} &= \smallsum  \lambda_{i}^{s} = \Tr{\mLambda^{s}} = \Tr{\mOmega_{11}^{s}} \nonumber \\
&= \Tr{\mOmega^{s}}  - \lambda^{s} = \Tr{(\mSigmat^{-1}\mSigma)^{s}}  - \lambda^{s}
\end{align}
Note that $c_{s}$ depends only on the eigenvalues of $\mSigmat^{-1}\mSigma$ and $\lambda$. The approximate Pearson's representation of the SNR loss writes
\begin{equation}\label{SNRloss_GER_approx_Pearson}
\lossger \approxdist \left[ 1+  \frac{\frac{c_{3}}{c_{2}} \chisquare{c_{3}^{2}/c_{2}^{3}}{0} + c_{1} - \frac{c_{2}^{2}}{c_{3}}}{\lambda \, \chisquare{2(K-N+2)}{0}} \right]^{-1}
\end{equation}
with $c_{s}$ given by \eqref{cs_GER}. Note that when $\lambda_{i}=1$ $\forall i$, then $c_{s}=N-1$, $c_{3}/c_{2}=1$, $c_{3}^{2}/c_{2}^{3}=2(N-1)$, $c_{1}-c_{2}^{2}/c_{3}=0$ and the approximation recovers the exact distribution, namely $\chisquare{2(N-1)}{0} / \chisquare{2(K-N+2)}{0}$. 

\begin{figure*}[tb]
	\centering
	\subfigure{
		\includegraphics[width=7.5cm]{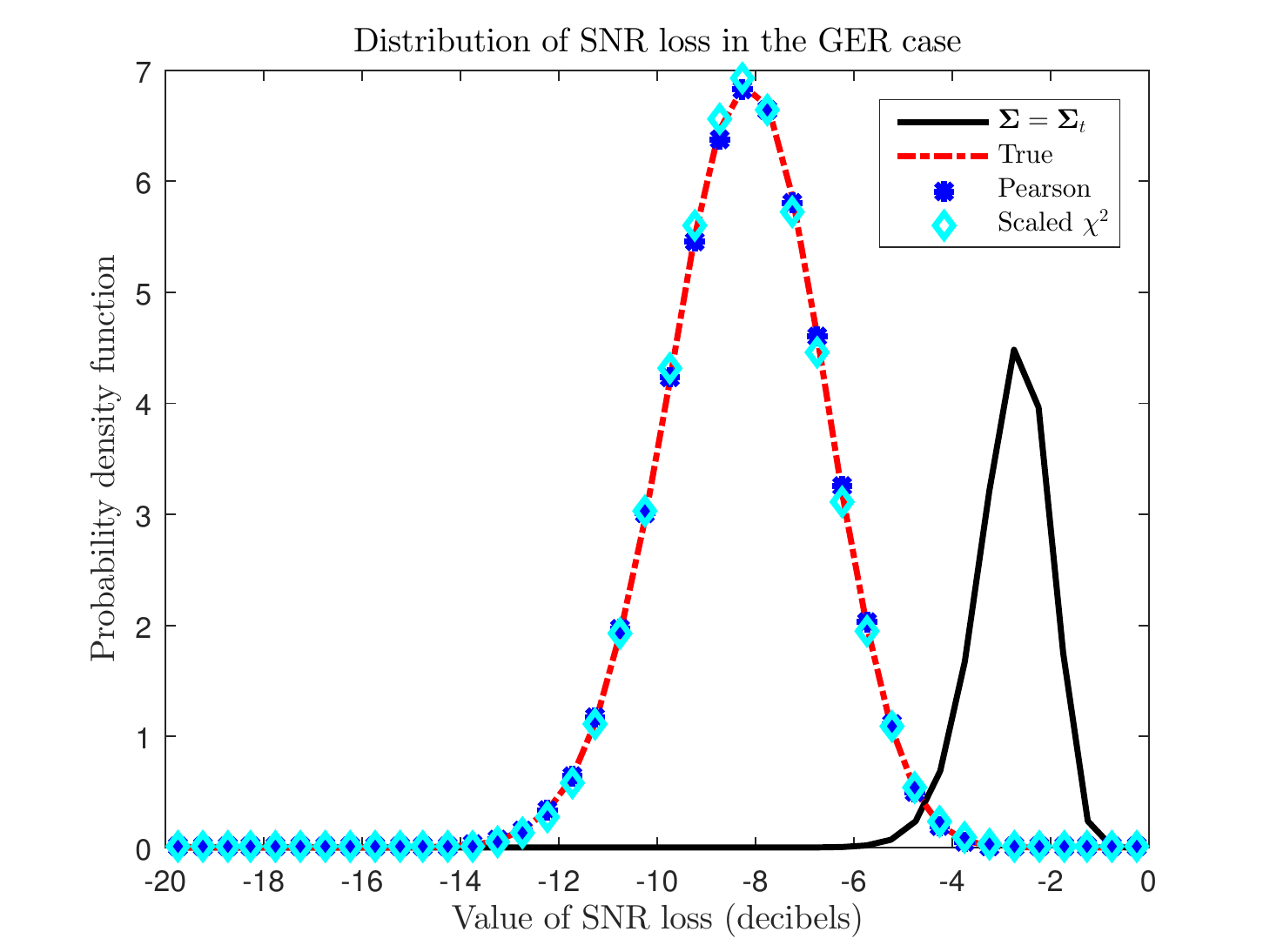}}
	\subfigure{%
		\includegraphics[width=7.5cm]{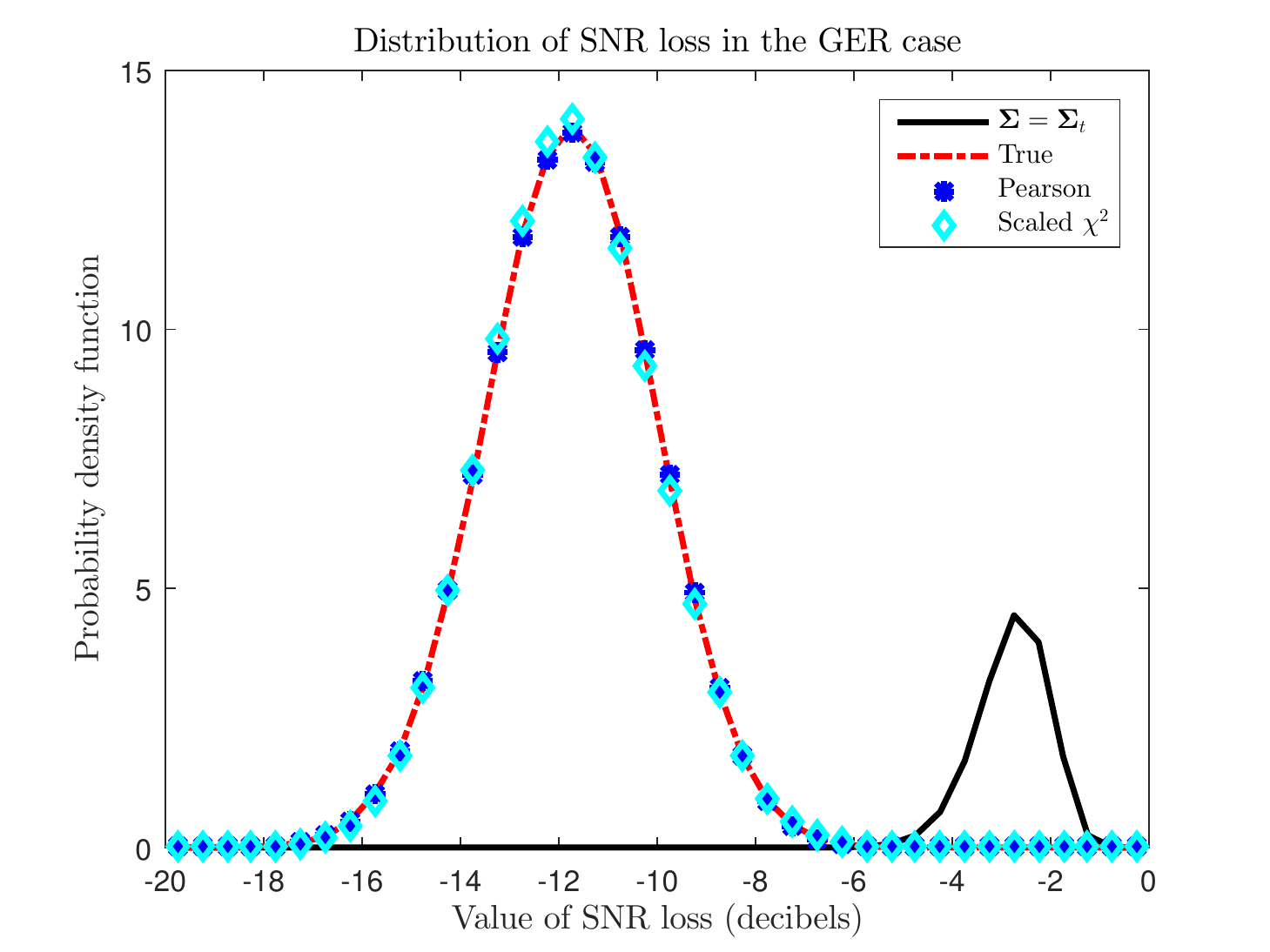}}\\
	\subfigure{%
		\includegraphics[width=7.5cm]{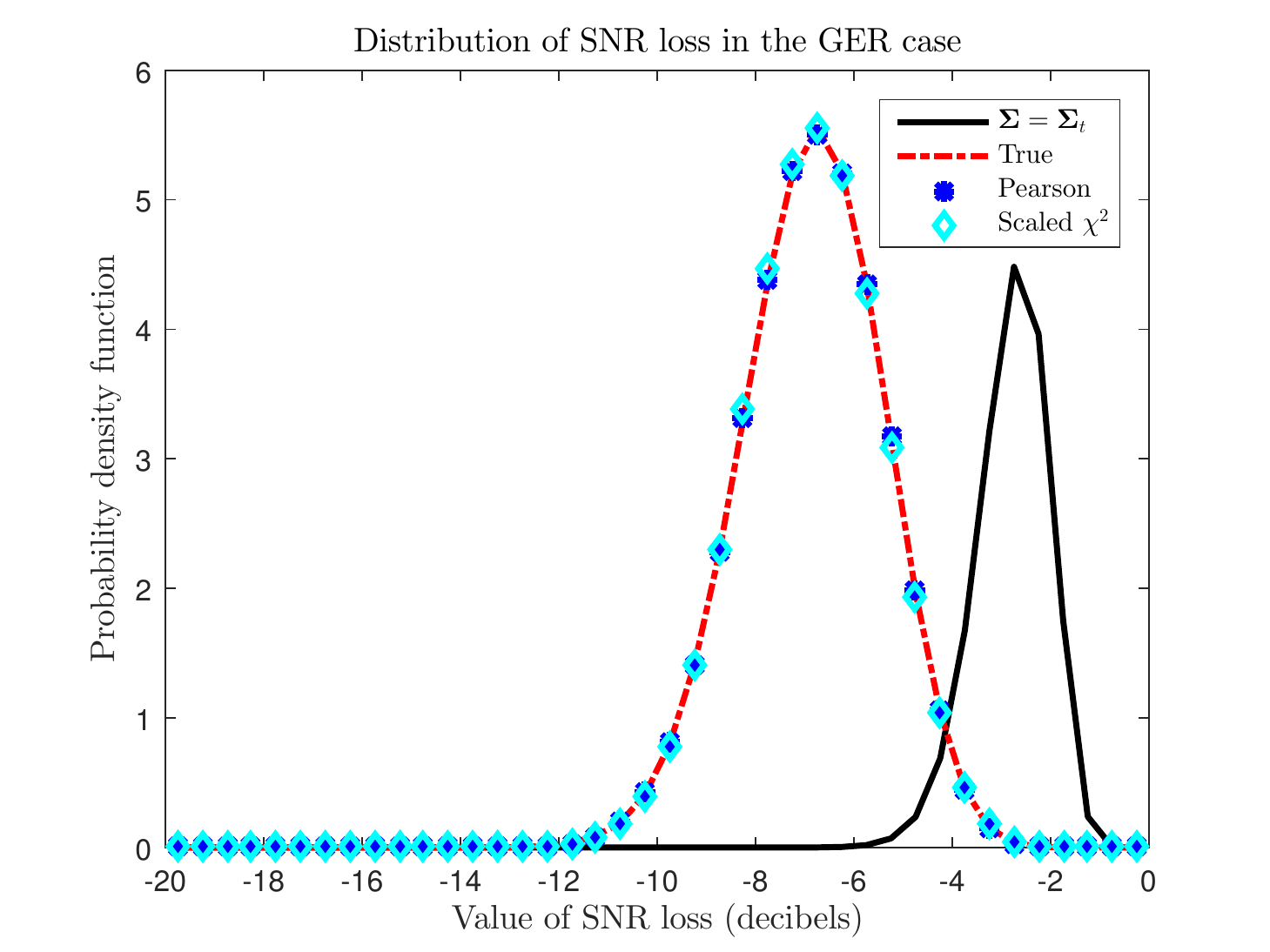}}
	\subfigure{%
		\includegraphics[width=7.5cm]{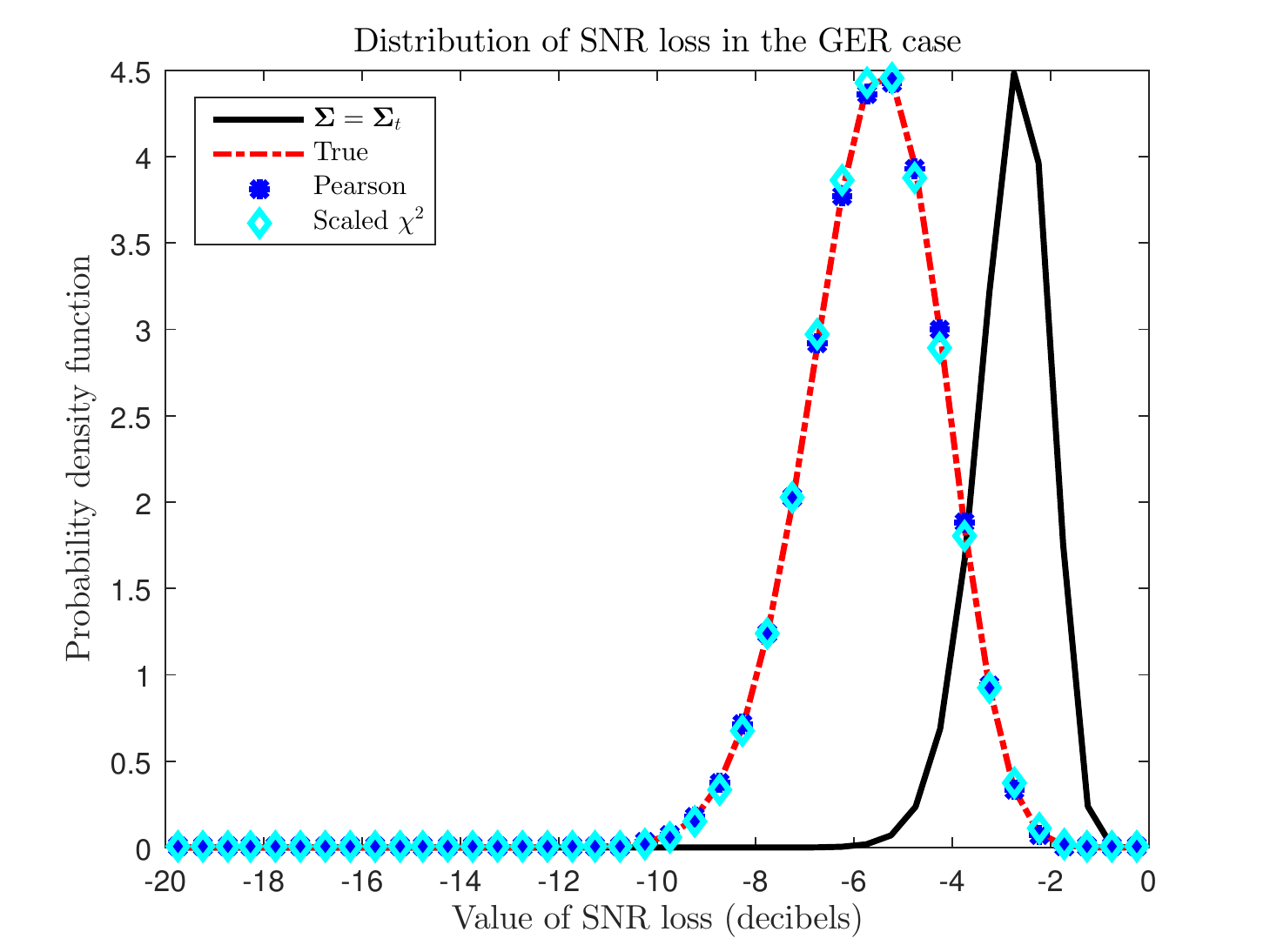}}
	\caption{True and approximated distribution of the SNR loss in the GER case under \eqref{GER_chol}. The mean values of $\W_{11}$ and $W_{22}$ are $\gamma \eye{N-1}$ and $\gamma$, and $10\log_{10}\gamma$ is uniformly distributed over [$-6$dB, $6$dB].}
	\label{fig:pdf_snrloss_GER=blockdiag_D_K=32}
\end{figure*}

While \eqref{SNRloss_GER_approx_Pearson} constitutes a reference and a simple representation, it does not allow for a closed-form expression of the p.d.f., mainly because some integral cannot be derived. Note also that we look for a simpler approximation, of the type \eqref{approx_representation_SNRloss}  which yields a p.d.f. of the form \eqref{approx_pdf_SNRloss}. Hence we look for an approximation $Q'_{a} = a \chisquare{\nu}{0}$ of $\smallsum_{i} \lambda_{i} \chisquare{h_{i}}{0}$. Since the distribution of $Q'_{a} $ contains only two parameters, only the two first moments of both variables can be made identical. It is straightforward to show that this yields $a=c_{2}/c_{1}$ and $\nu= c_{1}^{2}/c_{2}$. Therefore, we propose the following simpler approximation:
\begin{equation}\label{SNRloss_GER_approx_chi2}
\lossger \approxdist \left[ 1+  \frac{\frac{c_{2}}{\lambda c_{1}} \chisquare{c_{1}^{2}/c_{2}}{0}}{\chisquare{2(K-N+2)}{0}} \right]^{-1}
\end{equation}
Again, when $\lambda_{i}=1$ $\forall i$,  $c_{2}/c_{1}=1$, $c_{1}^{2}/c_{2}=2(N-1)$, and the exact distribution is recovered. 

We now evaluate the accuracy of \eqref{SNRloss_GER_approx_Pearson}-\eqref{SNRloss_GER_approx_chi2} in predicting the true distribution of the SNR loss, as represented in \eqref{SNRloss_GER}. Towards this end, we need to generate a matrix $\mSigmat$ which satisfies $\mSigmat^{-1}\vv = \lambda \mSigma^{-1}\vv $. Since the latter relation does not immediately provide a way to generate $\mSigmat$ from $\mSigma$, we show in the appendix that an equivalent and convenient way to satisfy the GER is that
\begin{equation}\label{GER_chol}
\Qv^{H} \mSigmat \Qv = \chol{\Qv^{H} \mSigma \Qv} \begin{pmatrix} \W_{11}^{-1} & \Mzero \\ \vzero & W_{22}^{-1} \end{pmatrix} \chol{\Qv^{H} \mSigma \Qv}
\end{equation}
where $\Qv=\begin{bmatrix} \Vorth & \vv \end{bmatrix}$. The formulation in \eqref{GER_chol} makes sense since it provides a partitioning of the matrices in the range spaces of $\Vorth$ and $\vv$. $\W_{11}$ [respectively $W_{22}$] quantifies the difference between $\mSigma$ and $\mSigmat$ in the subspace orthogonal to [respectively aligned with] $\vv$.

\begin{figure*}[tb]
	\centering
	\subfigure{
		\includegraphics[width=7.5cm]{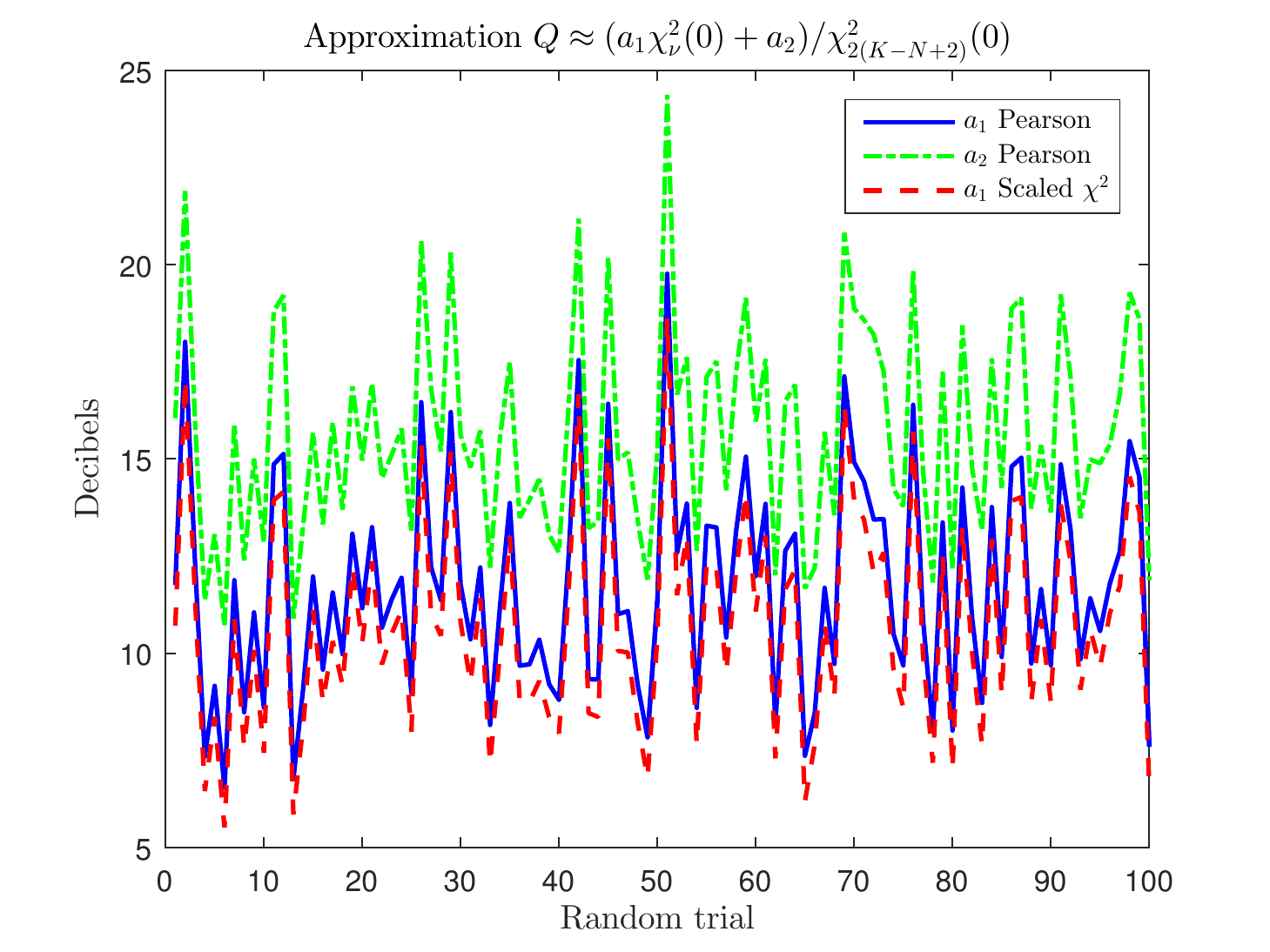}}
	\subfigure{%
		\includegraphics[width=7.5cm]{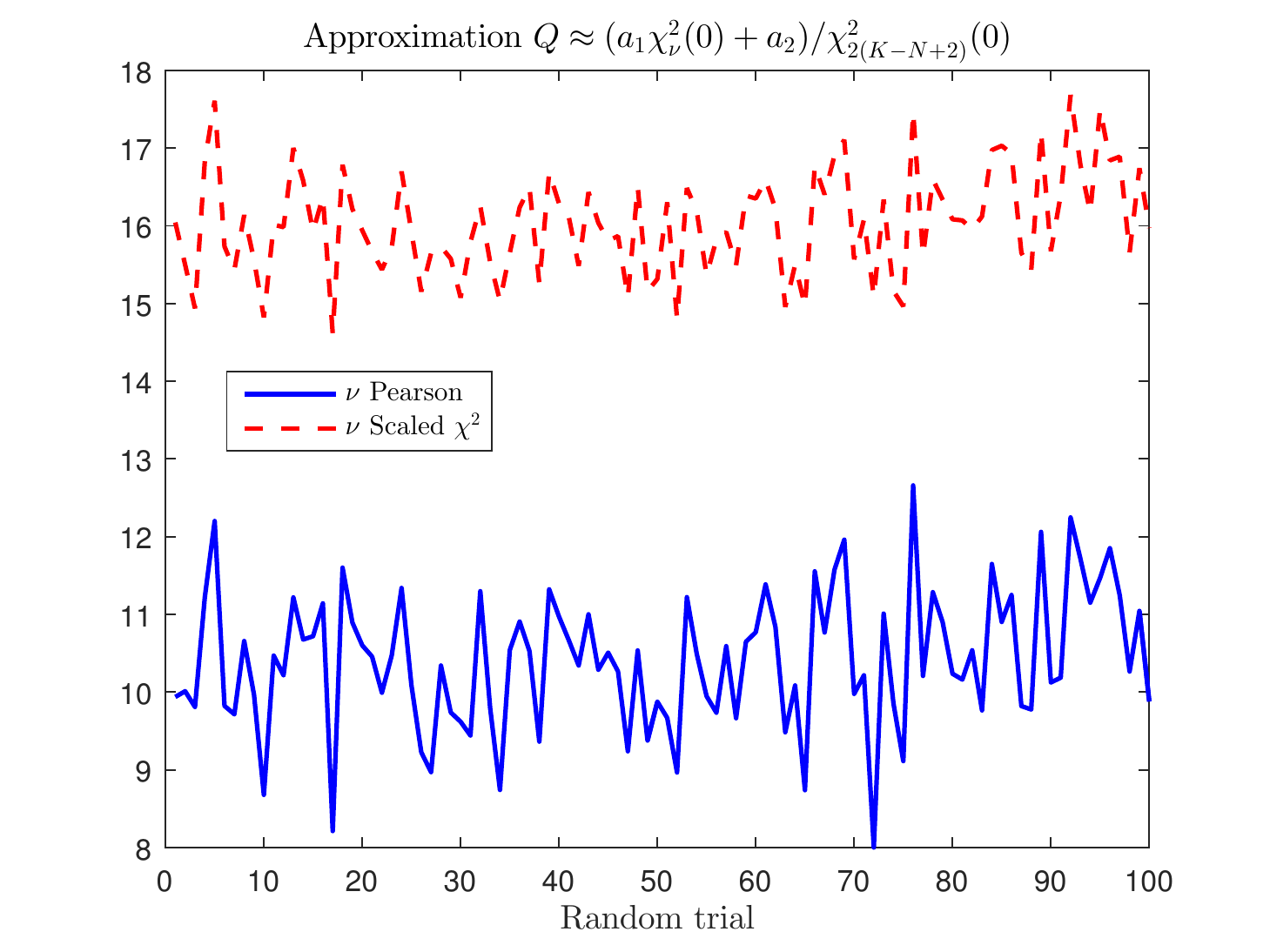}}
	\caption{Values of the parameters of the approximated (Pearson and scaled $\chi^{2}$) distributions of the SNR loss in the GER case under \eqref{GER_chol}. The mean values of $\W_{11}$ and $W_{22}$ are $\gamma \eye{N-1}$ and $\gamma$, and $10\log_{10}\gamma$ is uniformly distributed over [$-6$dB, $6$dB].}
	\label{fig:stats_approx_snrloss_chi2_GER=blockdiag_D_delta_mean_iW=6_K=32}
\end{figure*}

In order to assess the validity of the approximations \eqref{SNRloss_GER_approx_Pearson} and \eqref{SNRloss_GER_approx_chi2}, we generated $\mSigmat$ from $\mSigma$ as in \eqref{GER_chol} where $\W_{11}$ is drawn from a complex Wishart distribution and $W_{22}$ from a complex chi-square distribution. Both of them are such that $\E{\W_{11}^{-1}}=\gamma \eye{N-1}$ and $\E{W_{22}^{-1}}=\gamma$, and $\gamma$ is drawn randomly such that $10\log_{10}\gamma$ is uniformly distributed over [$-6$dB, $6$dB]. Figure \ref{fig:pdf_snrloss_GER=blockdiag_D_K=32} displays the true distribution of the SNR loss (estimated from $10^{6}$ Monte-Carlo trials) and its approximation either by Pearson's approach or by a scaled chi-square distribution, for $4$ different realizations of $\gamma$. As can be observed from this figure, both approximations are very accurate in predicting the actual distribution of the SNR loss. From these figures, one cannot distinguish the two and therefore the simpler approximation in \eqref{SNRloss_GER_approx_chi2} proves to be relevant. 
A second observation from this figure is that the distribution of the SNR loss can vary significantly when $\mSigmat$ changes. Finally, we notice that, due to covariance mismatch, the adaptive filter looses its ability to cancel interference and noise, resulting in a non negligible SNR loss.

It is also of interest to look at the values of the parameters of the approximated distributions. Towards this end, we generated $100$ random matrices as before and in Figure \ref{fig:stats_approx_snrloss_chi2_GER=blockdiag_D_delta_mean_iW=6_K=32} we display the values of the parameters of the two approximations. It is observed that the number of degrees of freedom in Pearson's approximation is usually smaller than that for the scaled $\chi^{2}$ distribution, and that it is somehow compensated by the coefficient $a_{2}$, which is non negligible. The parameters $a_{1}$ and $a$ are of the same order of magnitude, with $a$ usually a bit below $a_{1}$. Note also that the values of $\nu$ are quite stable over the realizations, while those of $a_{1}$, $a_{2}$ and $a$ vary more. To summarize, while the overall two distributions are quite similar, see Figure \ref{fig:pdf_snrloss_GER=blockdiag_D_K=32}, they are obtained with quite different parameters.

\section{Approximation of the distribution in the general case}
We now examine the general case where $\mSigma$ and $\mSigmat$ are arbitrary. As can be seen  from \eqref{SNRloss_vs_Omega} and \eqref{Q_Student} this constitutes a more challenging problem since we need now to approximate the distribution of a non-central quadratic form in $t$-distributed random vectors. Let us thus investigate the general problem of approximating $Q=V^{-1} \smallsum_{i} \lambda_{i} \chisquare{h_{i}}{V \delta_{i}}$ where $V \dist \chisquare{p}{0}$ by $Q_{a} = a \frac{Y}{W} \dist a \frac{\chisquare{\nu}{0}}{\chisquare{\mu}{0}}$. Towards this end, we will follow Pearson's approach and enforce that the mean, standard deviation and skewness of $Q$ and $Q_{a}$ be the same, or equivalently that their first three cumulants are identical. We first begin with deriving $\kappa_{n}(Q)$, $n=1,2,3$, then $\kappa_{n}(Q_{a})$, and finally we solve the system of equations $\kappa_{n}(Q)=\kappa_{n}(Q_{a})$.
\subsection{Derivation of $\kappa_{n}(Q)$, $n=1,2,3$}
In the sequel we derive the first three cumulants of $Q$. These derivations will rely on the fact that $\kappa_{n}(\smallsum_{i} \lambda_{i} \chisquare{h_{i}}{\delta_{i}} ) = 2^{n-1}(n-1)! \smallsum_{i} \lambda_{i}^{n} (h_{i}+n\delta_{i})$. From the definition of $Q$, we have
\begin{align}\label{kappa1(Q|V)}
\kappa_{1}(Q|V) &= \E{Q|V} = V^{-1} \smallsum_{i} \lambda_{i} (h_{i}+V\delta_{i}) \nonumber \\
&= \smallsum_{i}\lambda_{i}\delta_{i} + V^{-1} \smallsum_{i}\lambda_{i}h_{i}
\end{align}
and hence
\begin{equation}\label{kappa1(Q)}
\kappa_{1}(Q) = \E{\kappa_{1}(Q|V)} =  \smallsum_{i}\lambda_{i}\delta_{i} + \E{V^{-1}} \smallsum_{i}\lambda_{i}h_{i}
\end{equation}
In order to compute $\kappa_{2}(Q)=\mu_{2}(Q)$, we use the fact that
\begin{equation}\label{total_variance_Q}
\mu_{2}(Q) = \E{\mu_{2}(Q|V)} + \mu_{2}(\E{Q|V})
\end{equation}
However,
\begin{align}
\mu_{2}(Q|V) &= 2V^{-2} \smallsum_{i}\lambda_{i}^{2}(h_{i}+2V\delta_{i}) \nonumber \\
&= 2V^{-2} \smallsum_{i}\lambda_{i}^{2}h_{i} + 4V^{-1} \smallsum_{i}\lambda_{i}^{2}\delta_{i}
\end{align}
and thus
\begin{equation}
\E{\mu_{2}(Q|V)} = 2\E{V^{-2}} \smallsum_{i}\lambda_{i}^{2}h_{i} + 4\E{V^{-1}} \smallsum_{i}\lambda_{i}^{2}\delta_{i}
\end{equation}
Moreover
\begin{align}
\mu_{2}(\E{Q|V}) &= \mu_{2}(\smallsum_{i}\lambda_{i}\delta_{i} + V^{-1} \smallsum_{i}\lambda_{i}h_{i}) \nonumber \\
&= (\smallsum_{i}\lambda_{i}h_{i})^{2} \mu_{2}(V^{-1}) \nonumber \\
&= (\smallsum_{i}\lambda_{i}h_{i})^{2} \left[\E{V^{-2}}-\E{V^{-1}} ^{2} \right]
\end{align}
which finally yields
\begin{align}\label{kappa2(Q)}
\kappa_{2}(Q) &= \left[2\smallsum_{i}\lambda_{i}^{2}h_{i} + (\smallsum\lambda_{i}h_{i})^{2} \right] \E{V^{-2}} \nonumber \\
&+ 4\E{V^{-1}} \smallsum_{i}\lambda_{i}^{2}\delta_{i} - (\smallsum_{i}\lambda_{i}h_{i})^{2}\E{V^{-1}} ^{2}
\end{align}
It remains to obtain $\kappa_{3}(Q)=\mu_{3}(Q)$ which will be done using
\begin{equation}\label{total_cumulant_Q}
\mu_{3}(Q) = \E{\mu_{3}(Q|V)} + \mu_{3}(\E{Q|V}) + 3\cov{\E{Q|V},\mu_{2}(Q|V)}
\end{equation}
Now
\begin{align}
\mu_{3}(Q|V) &= 8V^{-3} \smallsum_{i}\lambda_{i}^{3}(h_{i}+3V\delta_{i}) \nonumber \\
&= 8V^{-3} \smallsum_{i}\lambda_{i}^{3}h_{i} + 24V^{-2} \smallsum_{i}\lambda_{i}^{3}\delta_{i}
\end{align}
and thus
\begin{equation}
\E{\mu_{3}(Q|V)} = 8\E{V^{-3}} \smallsum_{i}\lambda_{i}^{3}h_{i} + 24\E{V^{-2}} \smallsum_{i}\lambda_{i}^{3}\delta_{i}
\end{equation}
Next
\begin{align}
\mu_{3}(\E{Q|V}) &= \mu_{3}(\smallsum_{i}\lambda_{i}\delta_{i} + V^{-1} \smallsum_{i}\lambda_{i}h_{i}) \nonumber \\
&= (\smallsum_{i}\lambda_{i}h_{i})^{3} \mu_{3}(V^{-1}) \nonumber \\
&= (\smallsum_{i}\lambda_{i}h_{i})^{3} \left[\E{V^{-3}} -3\E{V^{-1}}\E{V^{-2}}+2\E{V^{-1}}^{3}\right]
\end{align}
\begin{align}
&\cov{\E{Q|V},\mu_{2}(Q|V)} \nonumber \\
&= \cov{V^{-1} \smallsum_{i}\lambda_{i}h_{i},2V^{-2} \smallsum_{i}\lambda_{i}^{2}h_{i} + 4V^{-1} \smallsum_{i}\lambda_{i}^{2}\delta_{i}} \nonumber \\
&= 2(\smallsum_{i}\lambda_{i}h_{i})(\smallsum_{i}\lambda_{i}^{2}h_{i}) \cov{V^{-1},V^{-2}} \nonumber \\
&+ 4(\smallsum_{i}\lambda_{i}h_{i})(\smallsum_{i}\lambda_{i}^{2}\delta_{i}) \cov{V^{-1},V^{-1}} \nonumber \\
&= 2(\smallsum_{i}\lambda_{i}h_{i})(\smallsum_{i}\lambda_{i}^{2}h_{i}) \left[\E{V^{-3}}-\E{V^{-1}}\E{V^{-2}}\right] \nonumber \\
&+ 4(\smallsum_{i}\lambda_{i}h_{i})(\smallsum_{i}\lambda_{i}^{2}\delta_{i}) \left[\E{V^{-2}}-\E{V^{-1}}^{2}\right]
\end{align}
We end up with
\begin{align}\label{kappa3(Q)}
\kappa_{3}(Q) &=\left[8 \smallsum_{i}\lambda_{i}^{3}h_{i} + (\smallsum_{i}\lambda_{i}h_{i})^{3} + 6(\smallsum_{i}\lambda_{i}h_{i})(\smallsum_{i}\lambda_{i}^{2}h_{i})\right]\E{V^{-3}} \nonumber \\
&+ \left[24 \smallsum_{i}\lambda_{i}^{3}\delta_{i} + 12(\smallsum_{i}\lambda_{i}h_{i})(\smallsum_{i}\lambda_{i}^{2}\delta_{i})\right]\E{V^{-2}} \nonumber \\
&-3\left[(\smallsum_{i}\lambda_{i}h_{i})^{3}  + 2(\smallsum_{i}\lambda_{i}h_{i})(\smallsum_{i}\lambda_{i}^{2}h_{i}) \right] \E{V^{-1}}\E{V^{-2}} \nonumber \\
&-12(\smallsum_{i}\lambda_{i}h_{i})(\smallsum_{i}\lambda_{i}^{2}\delta_{i})\E{V^{-1}}^{2} + 2 (\smallsum_{i}\lambda_{i}h_{i})^{3}\E{V^{-1}}^{3}
\end{align}
Equations \eqref{kappa1(Q)}, \eqref{kappa2(Q)} and \eqref{kappa3(Q)} enable to compute the first three cumulants of $Q$. Note that these formulas hold for any distribution of $V$. When $\V \dist \chisquare{2(K-N+2)}{0}$, one has
\begin{equation}
\E{V^{-k}} = \left( \prod_{i=1}^{k} 2(K-N+2-i)\right)^{-1}
\end{equation}
Therefore, $\E{V^{-3}}$ is finite only of $K > N+1$ and hence the approximation requires this condition to hold.
\subsection{Derivation of $\kappa_{n}(Q_{a})$, $n=1,2,3$}
We now compute the first three cumulants of $Q_{a}=aW^{-1}Y$ with $Y \dist \chisquare{\nu}{0}$ and $W \dist \chisquare{\mu}{0}$. We start with
\begin{equation}\label{kappa1(Qa)}
\kappa_{1}(Q_{a}) = \E{\E{Q_{a}|W}} = a\nu\E{ W^{-1}} 
\end{equation}
Next, since $\mu_{2}(Q_{a}|W)=a^{2}W^{-2}\mu_{2}(Y) = 2a^{2}\nu W^{-2}$, one has $\E{\mu_{2}(Q_{a}|W)}=2a^{2}\nu \E{W^{-2}}$. Furthermore, $\mu_{2}(\E{Q_{a}|W}) = (a\nu)^{2}\mu_{2}(W^{-1})$ and thus
\begin{align}\label{kappa2(Qa)}
\kappa_{2}(Q_{a}) &= 2a^{2}\nu\E{W^{-2}}  + (a\nu)^{2} \left[\E{W^{-2}} -  \E{W^{-1}}^{2}\right] \nonumber \\
&=a^{2} \nu (\nu + 2)\E{W^{-2}}  -  (a\nu)^{2}  \E{W^{-1}}^{2}
\end{align}
Also, $\mu_{3}(Q_{a}|W)=a^{3}W^{-3}\mu_{3}(Y) = 8a^{3}\nu W^{-3}$ which implies that $\E{\mu_{3}(Q_{a}|W)} = 8a^{3}\nu \E{W^{-3}}$. Additionally
\begin{align}
\mu_{3}(\E{Q_{a}|W}) &= (a\nu)^{3} \mu_{3}(W^{-1}) \nonumber \\
&= (a\nu)^{3} \left[\E{W^{-3}} - 3\E{W^{-1}}\E{W^{-2}}+2\E{W^{-1}}^{3}\right]
\end{align}
and
\begin{align}
\cov{\E{Q_{a}|W},\mu_{2}(Q_{a}|W)} &= \cov{a\nu W^{-1},2a^{2}\nu W^{-2}} \nonumber \\
&= 2a^{3}\nu^{2} \cov{W^{-1},W^{-2}}  \nonumber \\
&=2a^{3}\nu^{2} \left[\E{W^{-3}}-\E{W^{-1}}\E{W^{-2}}\right]
\end{align}
We end up with
\begin{align}\label{kappa3(Qa)}
\kappa_{3}(Q_{a}) &= a^{3}\nu (\nu+2)(\nu+4)\E{W^{-3}} \nonumber \\
&-3a^{3}\nu^{2}(\nu + 2)\E{W^{-1}}\E{W^{-2}} + 2(a\nu\E{W^{-1}})^{3}
\end{align}
Since $W \dist \chisquare{\mu}{0}$, $\E{W^{-k}} = \left( \prod_{i=1}^{k} (\mu-2i)\right)^{-1}$.

\begin{figure*}[tb]
	\centering
	\subfigure{%
		\includegraphics[width=7.5cm]{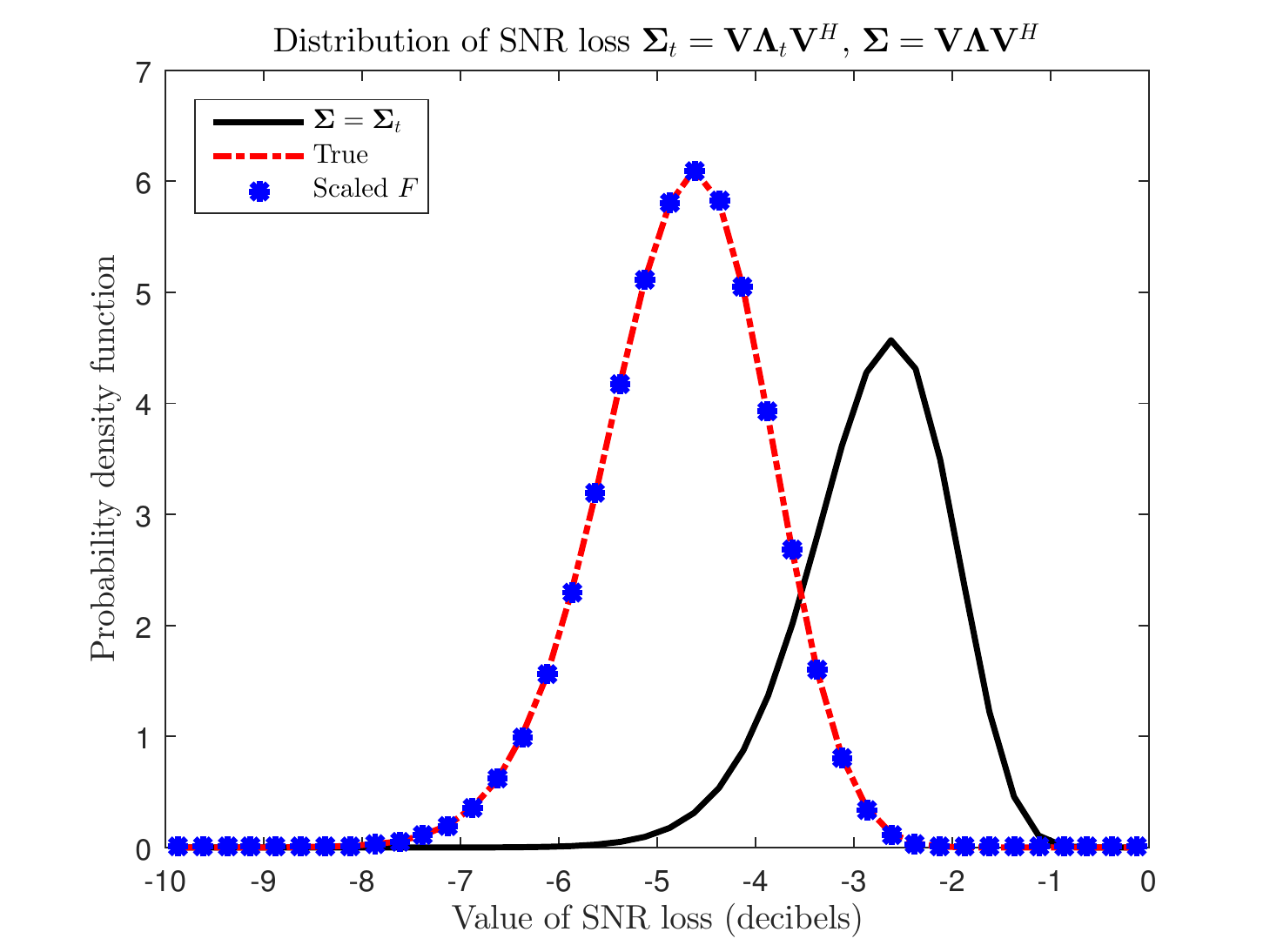}}	
	\subfigure{
		\includegraphics[width=7.5cm]{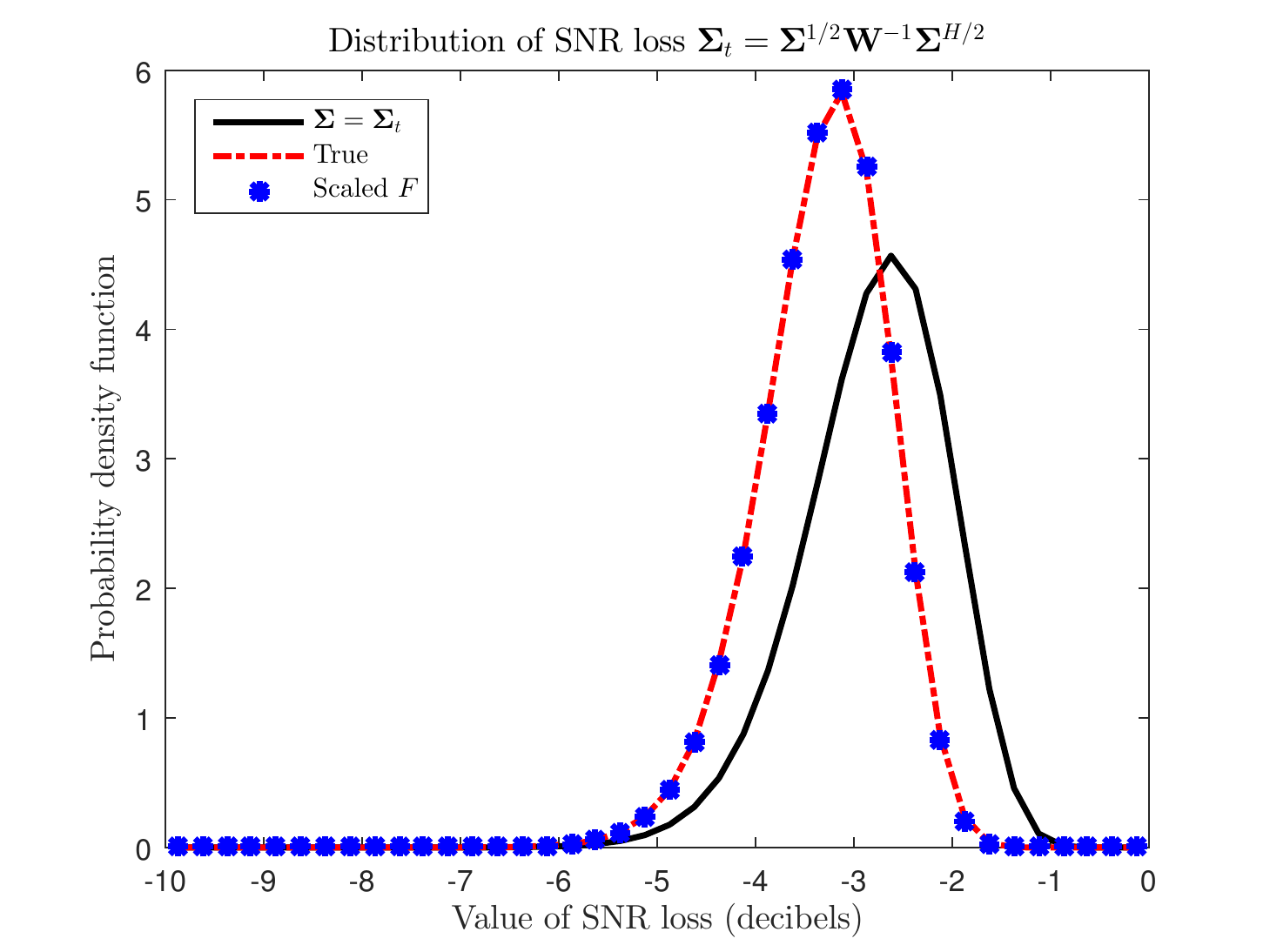}}\\
	\subfigure{%
		\includegraphics[width=7.5cm]{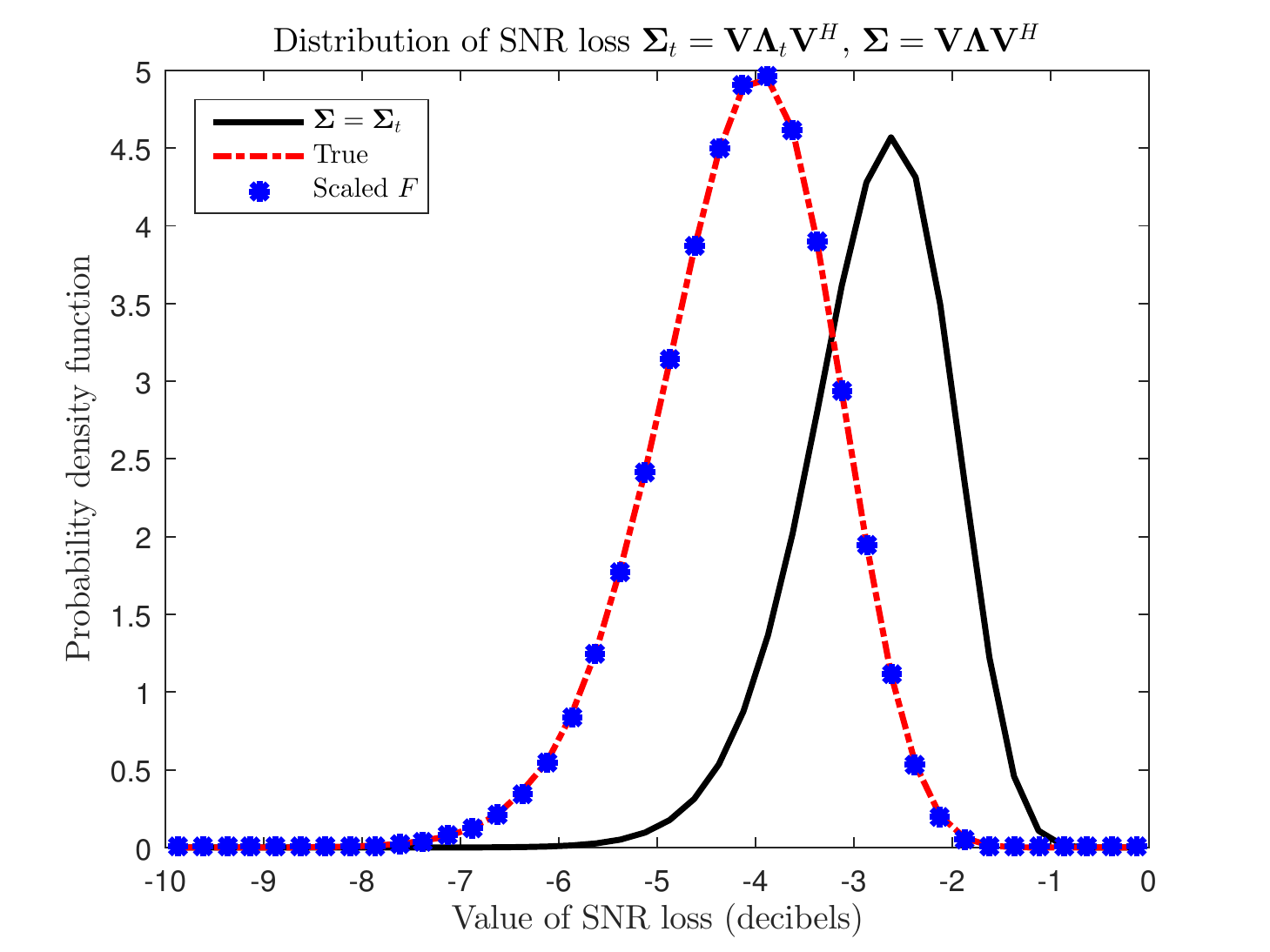}} 	
	\subfigure{%
		\includegraphics[width=7.5cm]{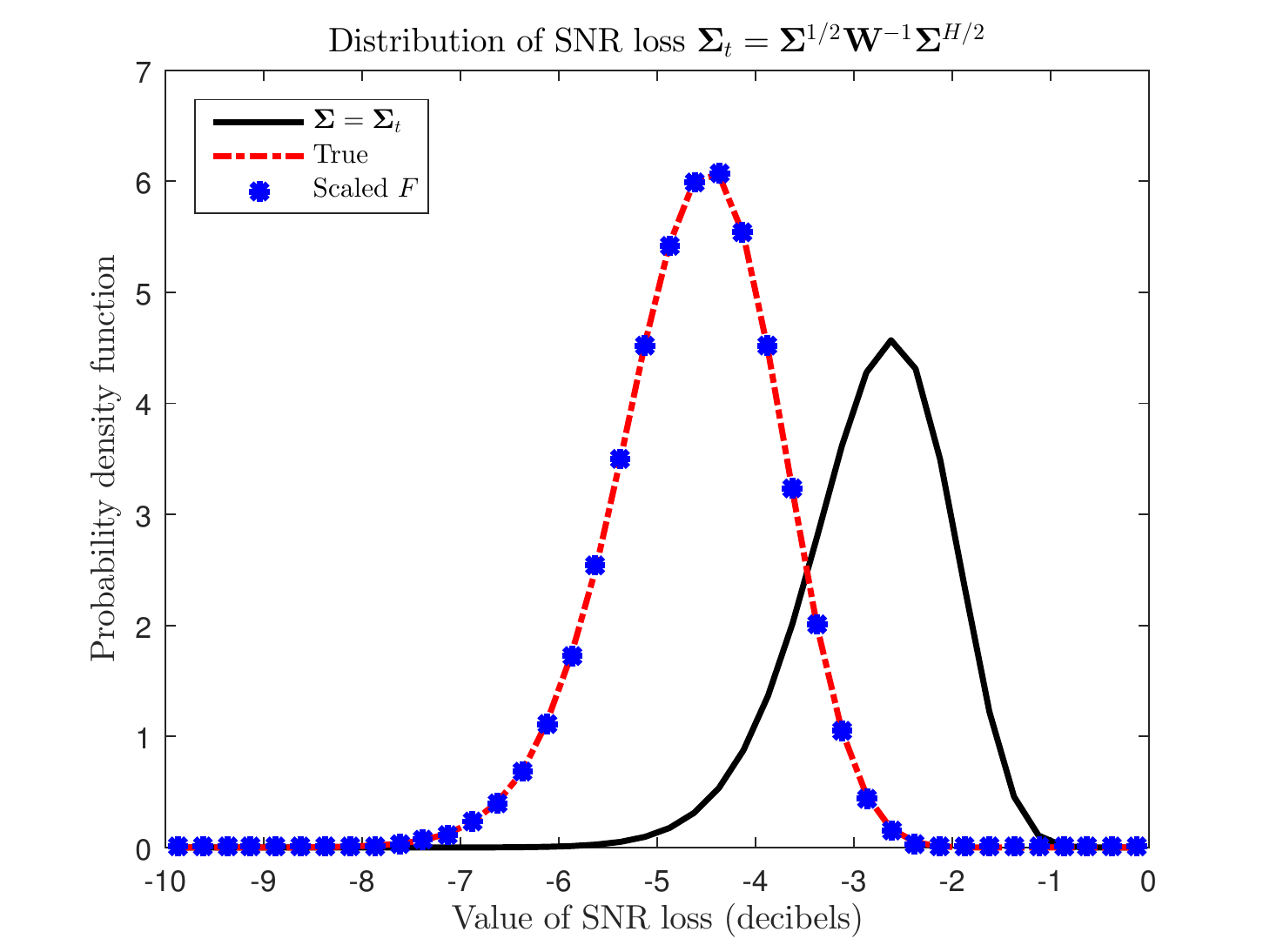}}
	\caption{True and approximated distribution of the SNR loss in the general case. Either  $\mSigmat$ and $\mSigma$ share the same eigenvectors but have different eigenvalues (left panel) or $\mSigmat=\mSigma^{1/2}\W^{-1}\mSigma^{H/2}$ with $\W$ Wishart distributed (right panel).}
	\label{fig:pdf_snrloss_scaledF_K=32}
\end{figure*}

\subsection{Finding the coefficients of the approximation}
We now find $a$, $\nu$ and $\mu$ such that $\kappa_{n}(Q_{a})  = \kappa_{n}(Q)$ for $n=1,2,3$. Let us temporarily omit the dependence on $Q$ and simply write $\kappa_{n}$. We note that $\E{W^{-1}}=(\mu-2)^ {-1}$,  $\E{W^{-2}}=\E{W^{-1}}(\mu-4)^ {-1}$ and $\E{W^{-3}}=\E{W^{-2}}(\mu-6)^ {-1}$.Writing $\kappa_{1}(Q_{a})  = \kappa_{1}(Q)=\kappa_{1}$ yields 
\begin{equation}\label{1st_equation}
(\mu-2) \kappa_{1} = a\nu
\end{equation}
For $n=2$, one can write
\begin{align}
\kappa_{2} &= a^{2} \nu (\nu + 2)\E{W^{-2}}  -  (a\nu)^{2}  \E{W^{-1}}^{2} \nonumber \\
&= \left[a\nu\E{W^{-1}}\right]\frac{a(\nu + 2)}{\mu-4} - \left(a\nu\E{W^{-1}}\right)^{2}
\end{align}
so that
\begin{equation}\label{2nd_equation}
(\mu-4) \left[\kappa_{2}+\kappa_{1}^{2}\right] = a(\nu + 2) \kappa_{1}
\end{equation}
As for $\kappa_{3}$ we have
\begin{align}
\kappa_{3} &= \left[a^{2} \nu (\nu + 2)\E{W^{-2}}\right] \frac{a(\nu + 4)}{\mu-6}\nonumber \\
&-3\left[a^{2} \nu (\nu + 2)\E{W^{-2}}\right]\left[a\nu\E{W^{-1}}\right] + 2\left(a\nu\E{W^{-1}}\right)^{3}
\end{align}
which can be rewritten as
\begin{equation}\label{3rd_equation}
(\mu-6) \left[\kappa_{3} + 3\kappa_{1}(\kappa_{2}+\kappa_{1}^{2}) - 2\kappa_{1}^{3}\right] = a(\nu + 4) \left[\kappa_{2}+\kappa_{1}^{2}\right]
\end{equation}
Equations \eqref{1st_equation}, \eqref{2nd_equation} and \eqref{3rd_equation} constitute indeed a \emph{linear system of equations} in the variables $\mu$, $a\nu$ and $a$. Indeed if we let $\veta = \begin{bmatrix} \mu & a\nu & a \end{bmatrix}^{T}$,  one needs to solve the system $\A\veta=\vb$ where
\begin{subequations}\label{A_eta=b}	
	\begin{align}
	\A &= \begin{pmatrix} \kappa_{1} & -1 & 0 \\ \kappa_{2}+\kappa_{1}^{2} & -\kappa_{1} & -2\kappa_{1} \\ \kappa_{3}+3\kappa_{1}\kappa_{2}+\kappa_{1}^{3} & -(\kappa_{2}+\kappa_{1}^{2}) & -4 (\kappa_{2}+\kappa_{1}^{2})  \end{pmatrix} \\ 
	\vb &= \begin{pmatrix} 2\kappa_{1} \\ 4(\kappa_{2}+\kappa_{1}^{2}) \\ 6(\kappa_{3}+3\kappa_{1}\kappa_{2}+\kappa_{1}^{3}) \end{pmatrix}
	\end{align}
\end{subequations}
The determinant of $\A$ is readily shown to be $\det{\A}=2(\kappa_{1}\kappa_{3}-2\kappa_{2}^{2})$ and therefore the solution is unique provided that $\kappa_{1}\kappa_{3} \neq 2\kappa_{2}^{2}$. 
This condition is in fact logical since, with the chosen approximation $Q_{a} \dist a \chisquare{\nu}{0} / \chisquare{\mu}{0}$, it is readily found that $\kappa_{1}(Q_{a})\kappa_{3}(Q_{a}) > 2\kappa^{2}_{2}(Q_{a})$. Therefore if $Q$ is such that $\kappa_{1}\kappa_{3} = 2\kappa^{2}_{2}$, then the system of equations $\kappa_{n}=\kappa_{n}(Q_{a})$, $n=1,2,3$ cannot have a solution. In this case one needs to find a different approximation for $Q$. Since the expression of these cumulants is rather complicated, we cannot prove that one always has $\kappa_{1}\kappa_{3}\neq 2\kappa_{2}^{2}$ (although in our simulations we did not encounter a case where this does not hold)  and hence we assume in the sequel that  this condition holds. Assuming that $\kappa_{1}\kappa_{3} \neq 2\kappa_{2}^{2}$, \eqref{A_eta=b} can be solved analytically, yielding the \emph{closed-form solution}:
\begin{subequations}\label{a_nu_mu}
	\begin{align}
	a &= \frac{\kappa_{2}\kappa_{3}+4\kappa_{1}\kappa_{2}^{2}-\kappa_{1}^{2}\kappa_{3}}{\kappa_{1}\kappa_{3}-2\kappa_{2}^{2}} \\
	\nu &= \frac{4\kappa_{1}(\kappa_{1}\kappa_{3}+\kappa_{1}^{2}\kappa_{2}-\kappa_{2}^{2})}{\kappa_{2}\kappa_{3}+4\kappa_{1}\kappa_{2}^{2}-\kappa_{1}^{2}\kappa_{3}} \\
	\mu &= 2 + 4 \frac{\kappa_{1}\kappa_{3}+\kappa_{1}^{2}\kappa_{2}-\kappa_{2}^{2}}{\kappa_{1}\kappa_{3}-2\kappa_{2}^{2}}
	\end{align}
\end{subequations}	
Therefore, the parameters of the approximation can be computed very easily once the cumulants of $Q$ are available. Note that this scaled $F$-distribution approximation holds whatever the distribution of $Q$ but of course derives from and is suitable for the case where $Q$ is a non-central quadratic form in $t$-distributed variables.

\begin{figure*}[tb]
	\centering
	\subfigure{%
		\includegraphics[width=7.5cm]{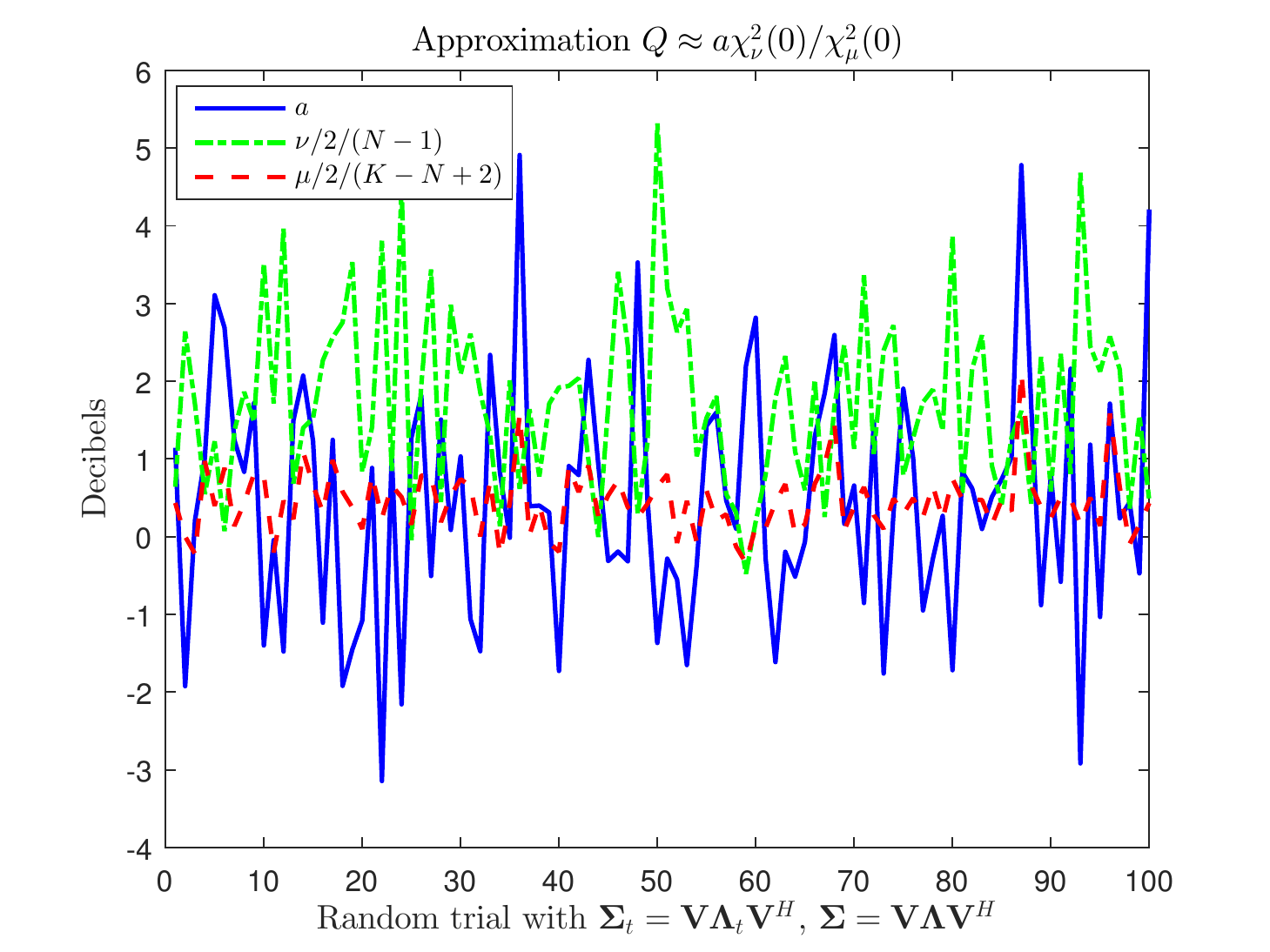}}
	\subfigure{
		\includegraphics[width=7.5cm]{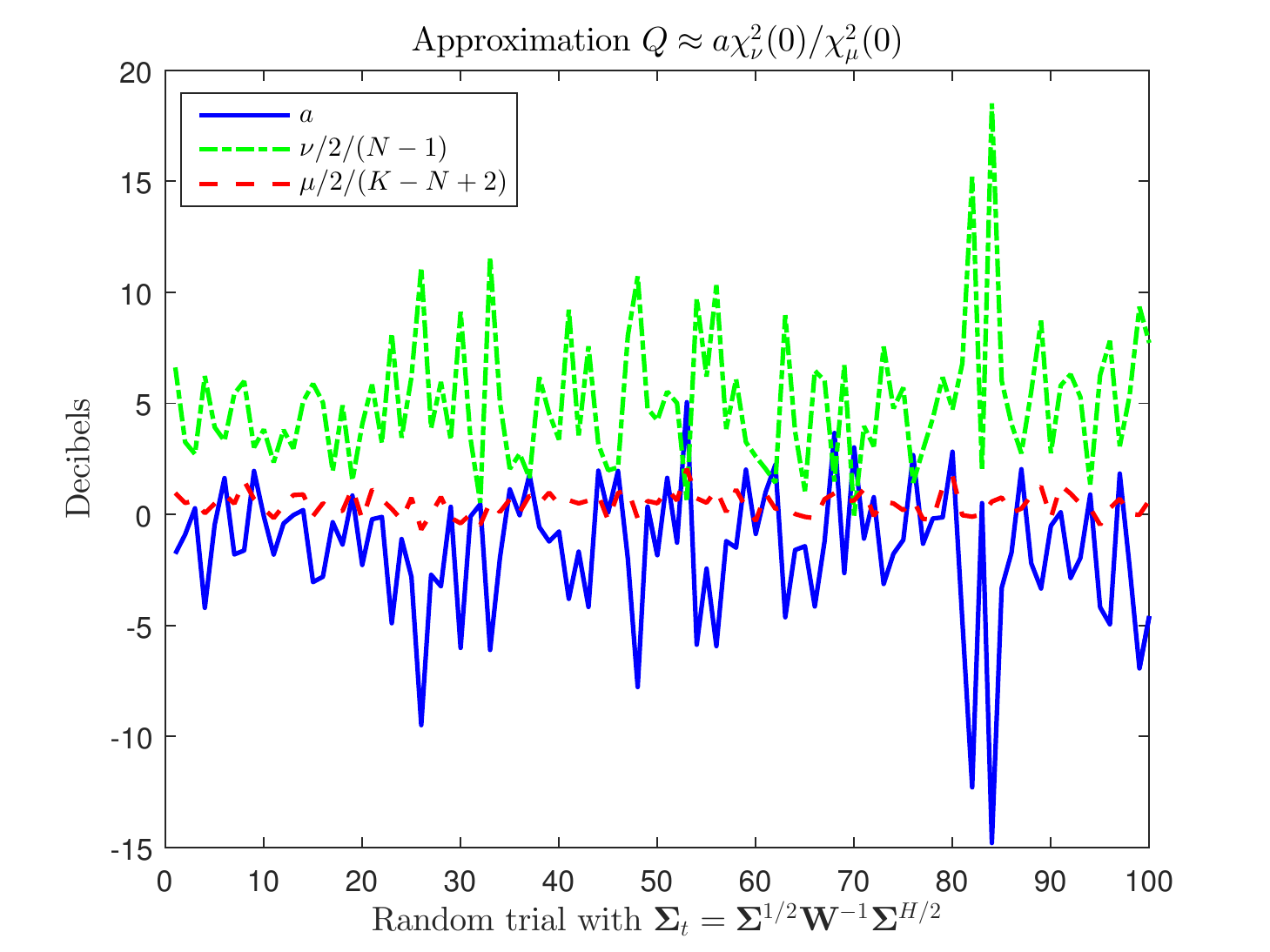}}
	\caption{Values of the parameters of the approximated scaled $F$-distributions of the SNR loss in the general case. Either  $\mSigmat$ and $\mSigma$ share the same eigenvectors but have different eigenvalues (left panel) or $\mSigmat=\mSigma^{1/2}\W^{-1}\mSigma^{H/2}$ with $\W$ Wishart distributed (right panel).}
	\label{fig:stats_approx_snrloss_scaledF_K=32}
\end{figure*}

\subsection{Numerical illustration}
We now evaluate whether this approximation of a non-central quadratic form in $t$-distributed variables is accurate. We consider here two types of mismatch: either  $\mSigmat$ and $\mSigma$ share the same eigenvectors but have different eigenvalues or $\mSigmat=\mSigma^{1/2}\W^{-1}\mSigma^{H/2}$. In the former case, the eigenvalues of $\mSigmat$ are generated randomly as $\lambda_{n}(\mSigmat) = \alpha_{n} \lambda_{n}(\mSigma)$ where $10\log_{10} \alpha_{n}$ is uniformly distributed over [$-6$dB, $6$dB]. In the latter case $\W$ is a Wishart matrix with mean value $\gamma \eye{N}$ and $10\log_{10}\gamma$ is uniformly distributed over [$-6$dB, $6$dB]. 

In Figure \ref{fig:pdf_snrloss_scaledF_K=32} we display the true distribution of $Q$ as well as its approximation for $2$ different realizations of $\gamma$ and $\alpha_{n}$. This figure clearly shows that the approximation is rather accurate as the approximated distribution fits very well the true distribution. Therefore  the approximation in \eqref{approx_representation_SNRloss} with $a$, $\nu$ and $\mu$ computed by \eqref{a_nu_mu} turns out to be effective.

Next, similarly to what was done previously, we investigate the	values of the parameters $a$, $\nu$ and $\mu$ of the approximation. The results are displayed in Figure \ref{fig:stats_approx_snrloss_scaledF_K=32} for $100$ different realizations of $\gamma$ and $\alpha_{n}$. This figure indicates that the value of $\mu$ -the number of degrees of freedom in the the chi-square distribution at the denominator- is quite stable while the values of the parameters of the chi-square distribution at the numerator, namely $a$ and $\nu$ vary more significantly, as if it was these two parameters that enable to adapt to the actual distribution.

\begin{figure}[htb]
	\centering
	\includegraphics[width=7.5cm]{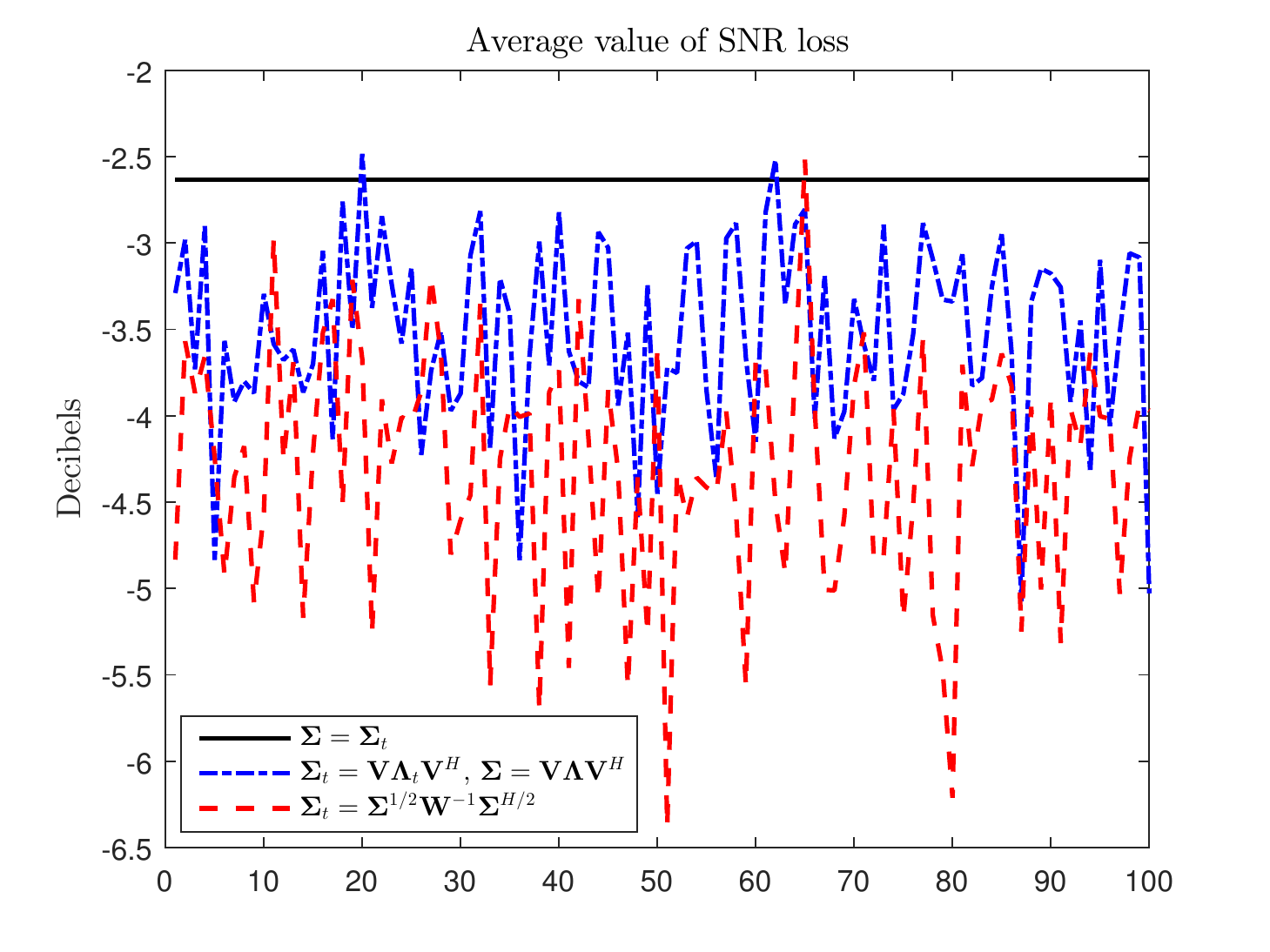}
	\caption{Mean value of the SNR loss in the general case.}
	\label{fig:mean_approx_snrloss_scaledF_K=32}
\end{figure}

The availability of a simple, easy-to-compute approximation makes it convenient to obtain, in a straightforward way, some relevant statistics about the SNR loss, for instance its mean value or the probability that it exceeds some threshold. As an illustration, for the $100$ random trials of the previous experiment, we computed, from the values of $a$, $\nu$ and $\mu$, the mean value of the SNR loss. The results are reported in Figure \ref{fig:mean_approx_snrloss_scaledF_K=32}. They provide another point of view compared to the p.d.f. and clearly mark the degradation induced by covariance mismatch since $\E{\loss}$ is well below its value in the case of no mismatch. It can also be observed that a covariance mismatch of the type $\mSigmat=\mSigma^{1/2}\W^{-1}\mSigma^{H/2}$ is more severe than a mismatch between the eigenvalues of $\mSigma$ and $\mSigmat$, at least in the case considered here.

\section{Conclusions}
In this paper we considered analyzing the distribution of the SNR loss when the adaptive filter is trained with covariance mismatched samples. We showed that the general representation of the SNR loss depends on a non-central quadratic form in $t$-distributed variables. When the generalized eigenrelation is satisfied, this simplifies to  a central quadratic form in normal variables. In the latter case we studied a Pearson-type approximation and, for both cases, a scaled $F$-distribution was proposed either using two or three moments. We provided closed-form expressions of the parameters of the approximated distributions and showed that they depend on the eigenvalues of $\mSigmat^{-1}\mSigma$ and on $(\vv^{H}\mSigmat^{-1}\vv)/(\vv^{H}\mSigma^{-1}\vv)$. A numerical study attested to the accuracy of the approximation in describing the actual distribution of the SNR loss. It enables one to have a simple, closed-form approximation of the form $\loss \approxdist \left[1 + a \chisquare{\nu}{0} / \chisquare{\mu}{0}\right]^{-1}$ and to assess rather quickly the impact of covariance mismatch.

\appendix
\section{Generation of matrices satisfying the GER \label{app:GER}}
In this appendix, we rewrite the GER in a form that enables an easy way to compute matrices satisfying this relation. Let $\Atilde$ and $\Btilde$ satisfying the GER. Let $\Qv=\begin{bmatrix} \Vorth & \vv \end{bmatrix}$, $\A=\Qv^{H}\Atilde\Qv$, $\B=\Qv^{H}\Btilde\Qv$  and note that
\begin{align}
\Atilde^{-1} \vv &= \lambda \Btilde^{-1} \vv \nonumber \\
&\Leftrightarrow (\Qv^{H}\Atilde\Qv)^{-1} \Qv^{H}\vv = \lambda (\Qv^{H}\Btilde\Qv)^{-1} \Qv^{H}\vv \nonumber \\
&\Leftrightarrow \A^{-1}\elast &= \lambda \B^{-1}\elast \nonumber \\
&\Leftrightarrow \begin{pmatrix} -\A_{11}^{-1} \A_{1 } A_{2.1}^{-1} \\  A_{2.1}^{-1} \end{pmatrix} = \lambda \begin{pmatrix} -\B_{11}^{-1} \B_{12} B_{2.1}^{-1} \\  B_{2.1}^{-1} \end{pmatrix}  \nonumber \\ &\Leftrightarrow \A_{11}^{-1} \A_{12} = \B_{11}^{-1} \B_{12} 
\end{align}
Let $\F=\chol{\A}$ and $\G=\chol{\B}$. Then 
\begin{align}
\A &= \begin{pmatrix} \F_{11} & \vzero \\ \F_{21} & F_{22} \end{pmatrix} \begin{pmatrix} \F_{11}^{H} & \F_{21}^{H} \\ \vzero & F_{22} \end{pmatrix} \nonumber \\
&= \begin{pmatrix} \F_{11}\F_{11}^{H} & \F_{11}\F_{21}^{H} \\ \F_{21}\F_{11}^{H} & \F_{21}\F_{21}^{H}+F_{22}^{2} \end{pmatrix}
\end{align}
which implies that $\A_{11}^{-1} \A_{12}=\F_{11}^{-H}\F_{21}^{H}$. So the GER is equivalent to $\F_{21}\F_{11}^{-1}=\G_{21}\G_{11}^{-1}$ and thus
\begin{align}
\G^{-1} \F &= \begin{pmatrix} \G_{11}^{-1} & \vzero \\ -G_{22}^{-1}\G_{21}\G_{11}^{-1} & G_{22}^{-1} \end{pmatrix} \begin{pmatrix} \F_{11} & \vzero \\ \F_{21} & F_{22} \end{pmatrix}  \nonumber \\
&= \begin{pmatrix} \G_{11}^{-1}\F_{11} & \vzero \\ -G_{22}^{-1}\G_{21}\G_{11}^{-1}\F_{11} + G_{22}^{-1}\F_{21} & G_{22}^{-1}F_{22} \end{pmatrix}  \nonumber \\
&= \begin{pmatrix} \G_{11}^{-1}\F_{11} & \vzero \\ \vzero & G_{22}^{-1}F_{22} \end{pmatrix} 
\end{align}
which is equivalent to
\begin{equation}
\G^{-1} \F \F^{H} \G^{-H} = \G^{-1} \A \G^{-H}  = \begin{pmatrix} \W_{11}^{-1} & \vzero \\ \vzero & W_{22}^{-1} \end{pmatrix} 
\end{equation}
Finally the GER can be recast as  $\A = \G \W^{-1} \G^{H}$ with $\W$ block-diagonal.

\end{document}